\documentclass[twocolumn,english,prx,reprint,superscriptaddress]{revtex4-1}
\usepackage[T2A,T1]{fontenc}
\usepackage[utf8]{inputenc}
\usepackage{verbatim}
\usepackage{amsmath}
\usepackage{amssymb}
\usepackage{graphicx}
\usepackage{color}
\usepackage{braket}

\usepackage{soul}
\setstcolor{red}

\makeatletter

\usepackage{graphicx} 
\graphicspath{{plots_submit/}} 

\usepackage[colorlinks,citecolor=blue,linkcolor=red,urlcolor=blue]{hyperref}

\usepackage{amsthm}
\usepackage{braket}

\makeatother

\usepackage{babel}

\begin{document}

\title{Topology and retardation effect of a giant atom in a topological waveguide}
\author{Weijun Cheng}
\affiliation{Center for Quantum Sciences and School of Physics, Northeast Normal University, Changchun 130024, China}
\author{Zhihai Wang}
\email{wangzh761@nenu.edu.cn}
\affiliation{Center for Quantum Sciences and School of Physics, Northeast Normal University, Changchun 130024, China}
\author{Yu-xi Liu}
\affiliation{School of Integrated Circuits, Tsinghua University, Beijing 100084, China}
\affiliation{Frontier Science Center for Quantum Information, Beijing 100084, China}

\begin{abstract}
The interaction between the quantum emitter and topological photonic system makes both the emitter and the photon behave in exotic ways. We here study a system that a giant atom is coupled to two points of a one-dimensional topological waveguide formed by the Su-Schrieffer-Heeger (SSH) chain. The topological nature of the hybrid system is studied. We find that the giant atom can act as an effective boundary and induce the chiral zero energy modes for the waveguide under the periodical boundary.  The properties of these modes are similar to those in the SSH model with open boundary. Meanwhile, the SSH waveguide, as a structured environment, induces the retarded effect and the non-Markovian dissipation of the giant atom. Our work may promote more studies on the interaction between matter and topological environment. Experimental demonstration for our study using superconducting quantum circuits is very possible within current technology.
\end{abstract}

\maketitle

\section{Introduction}
\label{Introduction}
Topological physics, which was originally explored in electronic systems~\cite{Kane,Qi}, has been promoted to photonic systems~\cite{TO} for engineering unconventional light behaviors. In the seminal work~\cite{Haldane}, the topological band structure was studied in photonic crystal with broken time-reversal symmetry. From then on, various photonic systems, e.g., the two-dimensional photonic crystal structures~\cite{ZW}, coupled waveguides or resonators~\cite{MC,MH,HZ,MP}, exciton-polaritons~\cite{PS}, metamaterials~\cite{WJ}, were proposed to study topological states of light, photonic analog to electronic quantum Hall effects, and synthetics of artificial gauge fields. These studies also further stimulate a new research direction for the interaction between topological photonic system and quantum emitters~\cite{JP,RJ,SB,MB,CA2019,LLx,DD,MA,Sd}, which are natural atoms and treated as point particles without considering their spatial shapes, and thus the quantum emitters are coupled to the topological photonic system via one point.

Simulations for topological physics using superconducting quantum circuits~\cite{Gu} have recently attracted extensive attentions. Topological transitions~\cite{PRoushan,MDSchroer,Tao}, invariants~\cite{Flurin}, and bands~\cite{X.Tan1,X.Tan2} were experimentally demonstrated using single superconducting qubits. Photonic localization and chiral ground-state currents using several superconducting qubits~\cite{P.Roushan1,P.Roushan} were experimentally observed. The topological waveguide was also constructed~\cite{WC,XG} via the Su-Schrieffer-Heeger (SSH)~\cite{WP} chain, which is characterized by non-zero winding number or zero-mode edge state(s) in the topological nontrivial phase with periodical or open boundary condition. The manipulation on topological states in SSH chain by quantized microwave field has further been studied~\cite{WN1,WN2}. Thus, the interaction between matter and topological photonic waveguide can be conveniently applied to study the coupling between the superconducting qubit and the topological microwave waveguide. However, in contrast to natural atoms, the superconducting qubits, acting as giant artificial atoms~\cite{AF,Gustafsson,Aref,Manenti,Noguchi,AFK1,Andersson,Bienfait,BK,LG,WZ,Xin,five}, can be engineered to be coupled with the waveguide via multiple points.  Such nonlocal  coupling with topological waveguide is not studied due to its highly interdisciplinary for superconducting quantum computation devices, topological physics, and quantum optics.

We here study the coupling between a giant artificial atom and a SSH chain waveguide via two coupling points under the periodical boundary condition
for the SSH chain. We study the exotic property of both the SSH chain and the giant atom.  On the one hand, we study the behaviors of the SSH chain. We find that the giant atom induced non-Bloch winding number~\cite{YSY} for the bulk state is the same as that of the bare SSH waveguide. Also, the giant atom results in the zero mode in the topologically non-trivial phase even that the topological waveguide is considered within the periodical boundary condition. This means that the giant atom acts as an effective boundary for the SSH chain. Our analytical results also show the chirality of the zero mode and the symmetry of the atom-waveguide dressed states. The zero mode is robust to the disorder and decoherence of the system. {On the other hand, the giant atom exhibits non-trivial dynamical behaviors in topological environment.}  The nonlocal two-point coupling between the giant atom and the topological waveguide results in the photon interference between the coupling points, which further induces the retardation effect and leads to non-Markovian dynamics.

\begin{figure}
  \centering
  \includegraphics[width=0.9\columnwidth]{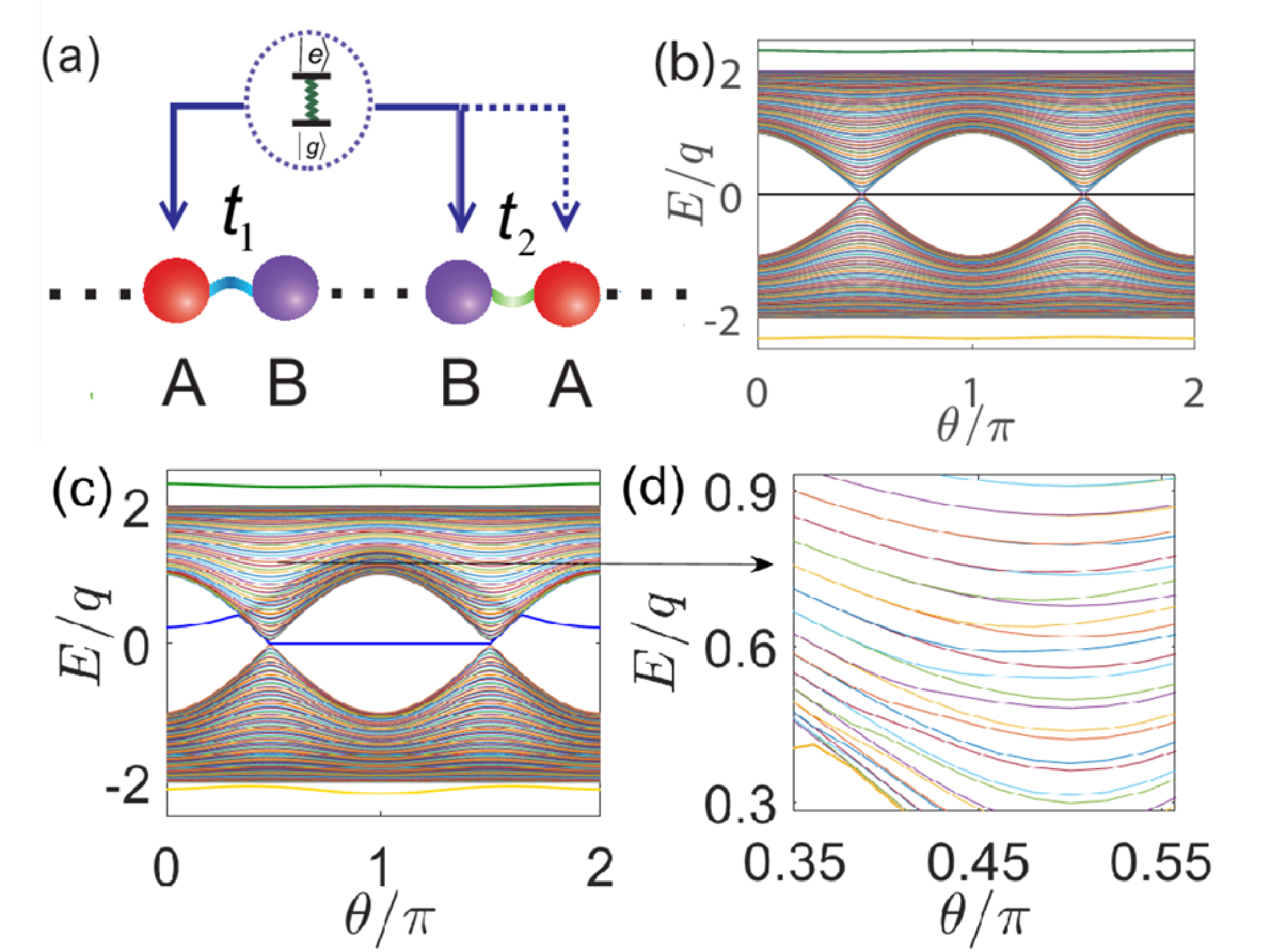}
  \caption{(a) Schematics of the SSH chain coupled to a giant atom via either A-A coupling (dashed) or A-B coupling (solid).  The energy spectrum versus $\theta$ in (b) for A-A coupling where the two-level is coupled to the waveguide via two $A$-type resonators,  and in (c) for A-B coupling under the periodical boundary condition. (d) The enlarged zoom in (c). The parameters are set as: $L=100$, $n=50$, $m=51$, $\omega=\Omega$, $\delta=0.5 q$ and $g=q$.}
\label{Energylevel}
\end{figure}

\section{Theoretical Model and Energy Spectrum}
\label{Theoretical model and energy spectrum}

\subsection{Theoretical Model}
\label{Theoretical model}
As schematically shown in Fig.~\ref{Energylevel} (a), we consider that a two-level giant atom, with ground state $|g\rangle$ and excited state $|e\rangle$, is coupled to a SSH chain waveguide via two points. The Hamiltonian of the SSH chain, consisting of $L$ unit cells, is
\begin{equation}\label{Eq:1}
H_{\rm{SSH}}=\sum_{l=1}^{L}[t_1\hat{C}^{\dag}_{A,l}\hat{C}_{B,l}+t_2\hat{C}^{\dag}_{A,l+1}\hat{C}_{B,l}]
+\rm{H.c.},
\end{equation}
in the rotating reference frame with $\omega$. Here, we assume that  each unit cell hosts $A$ and $B$ two resonators with the identical frequency $\omega$,  $\hat{C}_{A(B),l}$ is the annihilation operator of the point A (B) in the $l$th unit cell.
The parameters $t_1=q(1+\delta \cos\theta)$ and $t_2=q(1-\delta \cos\theta)$ are intracell and intercell couplings with $\delta$ being the dimerization strength. The parameter $\theta$ can vary from 0 to $2\pi$ continuously. Applying Fourier transform $\hat{C}_{\alpha,l}=1/\sqrt{L}\Sigma_k e^{ikl}\hat{C}_{\alpha,k}$ with $\alpha=A$ or $B$ to Eq.~(\ref{Eq:1}) under the periodical boundary condition, we can obtain the energy spectrum $E_{k\pm}=\pm \omega_k$ with $\omega_k=\sqrt{t_1^2+t_2^2+2t_1t_2\cos(k)}$, which is two-fold degenerate with $\omega_k=\omega_{-k}$. The topological property of the system is characterized by the winding number $\gamma_{\pm}$. As shown in Appendix~\ref{Su-Schrieffer-Heeger model}, $\gamma_{\pm}=0$ for $t_1>t_2$ ($0\leq\theta<\pi/2$ and $3\pi/2<\theta\leq 2\pi$) and $\gamma_{\pm}=1$ for $t_1<t_2$ ($\pi/2<\theta<3\pi/2$), correspond to topologically trivial and non-trivial phases, respectively.
 The energy spectrum for open boundary is given in Appendix~\ref{Su-Schrieffer-Heeger model} for odd and even sublattices, respectively. The zero-mode edge states are the topological signature of the waveguide.

The interaction Hamiltonian between the giant atom and the topological waveguide for the A-A (or A-B) coupling as shown in Fig.~\ref{Energylevel}(a) is given as $H_{{\rm AA}}=H_{\rm {SSH}}+H_{I,{\rm AA}}$ (or $H_{{\rm AB}}=H_{\rm {SSH}}+H_{I,{\rm AB}}$) with
\begin{subequations}
\begin{eqnarray}
H_{I,{\rm AA}}&=&\Omega|e\rangle\langle e|+g\sigma_+(C_{A,n}+C_{A,m})+\rm{H.c.},\\
H_{I,{\rm AB}}&=&\Omega|e\rangle\langle e|+g\sigma_+(C_{A,n}+C_{B,m})+\rm{H.c.}.
 \label{HI}
\end{eqnarray}
\end{subequations}
Here, $\Omega$ is the transition frequency of the two-level giant atom. In the following discussion of energy spectrum, we set $\Omega=\omega=0$. $g$ is the atom-waveguide coupling strength, $\sigma_+$ is the raising operator of the giant atom. $n$ and $m$ denote the positions of the unit cells. Without loss of generality, hereafter we assume $n<m$.

In the single-excitation subspace, the eigen wave-function of the system can be expressed as
\begin{equation}\label{psi}
|\psi\rangle=U_e|e,G\rangle+\sum_k A_k\hat{C}_{A,k}^{\dag}|g,G\rangle+\sum_k B_k\hat{C}_{B,k}^{\dag}|g,G\rangle,
\end{equation}
where $|G\rangle$ is the ground state of the SSH chain. Using the Schr\"{o}dinger equation $H|\psi\rangle = E|\psi\rangle$, we can obtain the eigen-energy $E$ corresponding to $|\psi\rangle$.

\subsection{Energy Spectrum Equations}
\label{Energy spectrum equations}
For A-A coupling, the eigen-energy $E$ is given as (see Appendix~\ref{Energy equation})
\begin{eqnarray}\label{eq:4}
E=\frac{g^{2}}{\pi}\int dk\left[\frac{E(1+\cos[k(m-n)])}{E^{2}-\omega_{k}^{2}}\right],
\end{eqnarray}
which implies that $E=0$ is always a solution. Moreover, for any state with the eigen-energy $E$, there is a partner state with  $-E$. This particle-hole symmetry~\cite{SR, LL} is shown in Fig.~\ref{Energylevel}(b), where the energy spectrum is plotted as a function of $\theta$. For A-B coupling, the eigen-energy satisfies (see Appendix~\ref{Energy equation})
\begin{eqnarray}\label{eq:5}
E=\frac{g^{2}}{\pi}\int dk\left[\frac{E+Q(k)}{E^{2}-\omega_k ^{2}}\right],
\end{eqnarray}
with
\begin{eqnarray}\label{Qk}
Q(k)=t_1\cos[k(m-n)]+t_2\cos[k(n-m-1)].
\end{eqnarray}

{To discuss the existence of the zero-mode, we set $E=0$,  the right side of Eq.~\eqref{eq:5} can be simplified to
\begin{eqnarray}
\frac{g^2}{\pi}\int_{-\pi}^{\pi}dk\frac{Q(k)}{-\omega_k^2}
=\begin{cases}-\frac{2g^2}{t_1}(-\frac{t_2}{t_1})^{m-n}& (t_1>t_2)\\
 0  & (t_2\geq t_1)\end{cases}.
 \label{ABequation2}
\end{eqnarray}}
Therefore, Eq.~(\ref{eq:5}) demonstrates that $E=0$ is the solution only for $t_2>t_1$, which is in the topological nontrivial phase ({$\pi/2<\theta< 3\pi/2$}) of the waveguide. However, for the state with the nonzero eigen-energy $E$, we cannot find a partner state with the eigen-energy $-E$. That is, the particle-hole symmetry is broken for the A-B coupling.
This has also been numerically verified in Fig.~\ref{Energylevel} (c). Moreover, the giant atom breaks the translation symmetry of the SSH waveguide, there are bound states outside the energy bands.

\begin{figure}\centering
\includegraphics[width=1 \columnwidth]{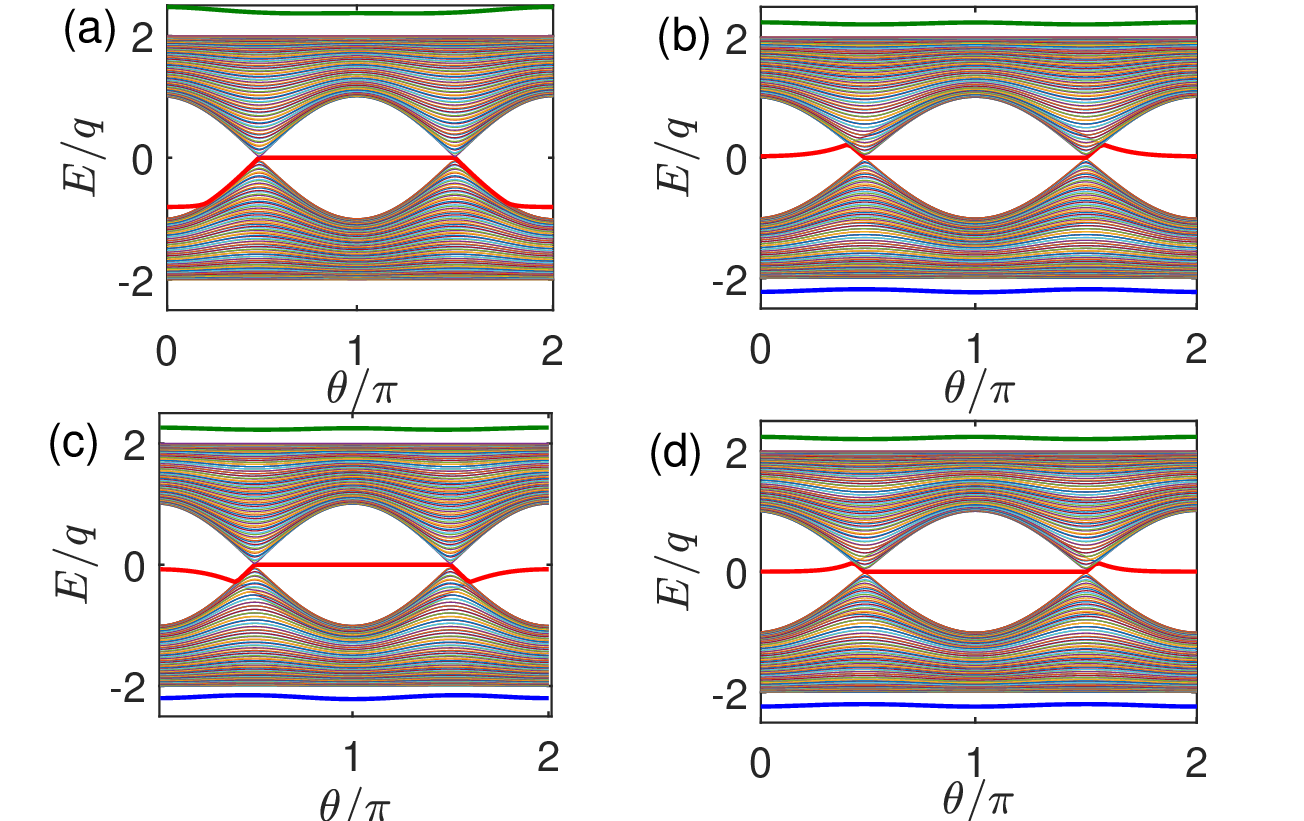}
\caption{The energy level diagram for A-B coupling setup. The parameters are set as: $L=100$, $\delta=0.5$ and $g=q$ and $d=|m-n|=0,3,2,5$ for (a), (b), (c), and (d), respectively.}
\label{ABlevel}
\end{figure}

As shown in Fig.~\ref{Energylevel}(c), we find that there is still an energy level in the gap for the A-B coupling.{ The nonzero energy level $E^g$ in the gap (in topological trivial phase) for the A-B coupling can be tuned on demand via adjusting the size of the giant atom, that is, the values of $n$ and $m$. In Fig.~\ref{ABlevel}, we show the energy level diagram for different $d=|m-n|$. For even $d$, as shown in Figs.~\ref{ABlevel}(a) and (c), $E^g$ is always negative, which is otherwise positive for odd $d$ as shown in Figs.~\ref{ABlevel}(b) and (d).  Moreover, as the increase of $d$, the value of $|E^g|$ approaches zero gradually. This can be understood from the expression of $Q$. Because $Q$ is a oscillation function of the wave vector $k$, and $d$ characterizes the oscillation frequency. Therefore, a large $d$ will naturally lead to a fast oscillation, and the contribution to $Q$ can be neglected via integration. As a result, the eigen energy approaches to zero.}

{Specially, from Eq.~\eqref{ABequation2}, we obtain $(-t_2/t_1)^{m-n}\rightarrow0$ when $(n-m)\rightarrow +\infty$.
Therefore, in this case, we predict that the eigen energy approaches to zero even in topologically trivial phase for the A-B coupling.
To understand it, we can regard the system as a combination of two open waveguides which are separated by the giant atom. The energy spectrum of the system is governed by the collective effect of  two open waveguides. When $(n-m)\rightarrow +\infty$, two open waveguides with the A-B and B-A open boundaries will lead to the existence of the zero-mode in $\theta\in(\pi/2,3\pi/2)$ and $\theta\in(0,\pi/2)\cup(3\pi/2,2\pi)$, respectively. As a result, the zero-mode exists in arbitrary phase.}

 \subsection{Non-trivial Degeneracy  Lifted by Giant Atom}
\label{Non-trivial degeneracy  lifted}

{In the Sec.~\ref{Energy spectrum equations}, we have shown that the two-fold degeneracy of the energy levels starts to be lifted in Fig.~\ref{Energylevel}(d).}
{To understand this non-trivial phenomenon in the A-B coupling setup, we rewrite the Hamiltonian in the momentum space as
\begin{eqnarray}
H_{\rm SSH}&=&\sum_k [E_{k+}|E_{k+}\rangle\langle E_{k+}|+E_{k-}|E_{k-}\rangle\langle E_{k-}|],\nonumber \\
H_{I,AB}&=&\sum_{k,\sigma=\pm}[g_{k\sigma}|G\rangle\langle E_{k\sigma}|\sigma^++{\rm H.c}],\nonumber \\
\end{eqnarray}
where
\begin{eqnarray}
g_{k\pm}=\frac{g}{\sqrt{2L}}(\frac{\omega_ke^{ikn}}{t_1+t_2e^{ik}}\pm e^{ikm}).
\end{eqnarray}
We recall that the frequencies of the atom and the bare resonator are set to be zero, the atomic excited state is separated from the energy levels inside the bands by the energy gap $E_{k\pm}\gg g$, and therefore can be adiabatically eliminated by use of the Schrieffer-Wolff transformation (also called Fr\"{o}lich-Nakajima transformation)~\cite{SW,SDD,HF,SN,YLI}.  Keeping up to the second order of atom-waveguide coupling $g$, we hence obtain the virtual coupling between the different energy levels of the SSH waveguide as
\begin{eqnarray}
\begin{split}
H_{\rm{eff}}&=\sum_{k,\sigma=\pm}E_{k\sigma}|E_{k\sigma}\rangle\langle E_{k\sigma}| \\
&+\sum_{k,k',(\sigma,\sigma')=\pm}G_{k\sigma,k'\sigma'}|E_{k\sigma}\rangle\langle E_{k'\sigma'}|,
\end{split}
\end{eqnarray}
where the effective coupling strength is explicitly given in Appendix~\ref{Non-trivial degeneracy  broken}. It shows that the
giant atom will induce the virtual coupling between the bulk modes of the SSH waveguide with same and different wave vectors, both inside the same band or different bands.}

\begin{figure*}
  \centering
  \includegraphics[width=15cm]{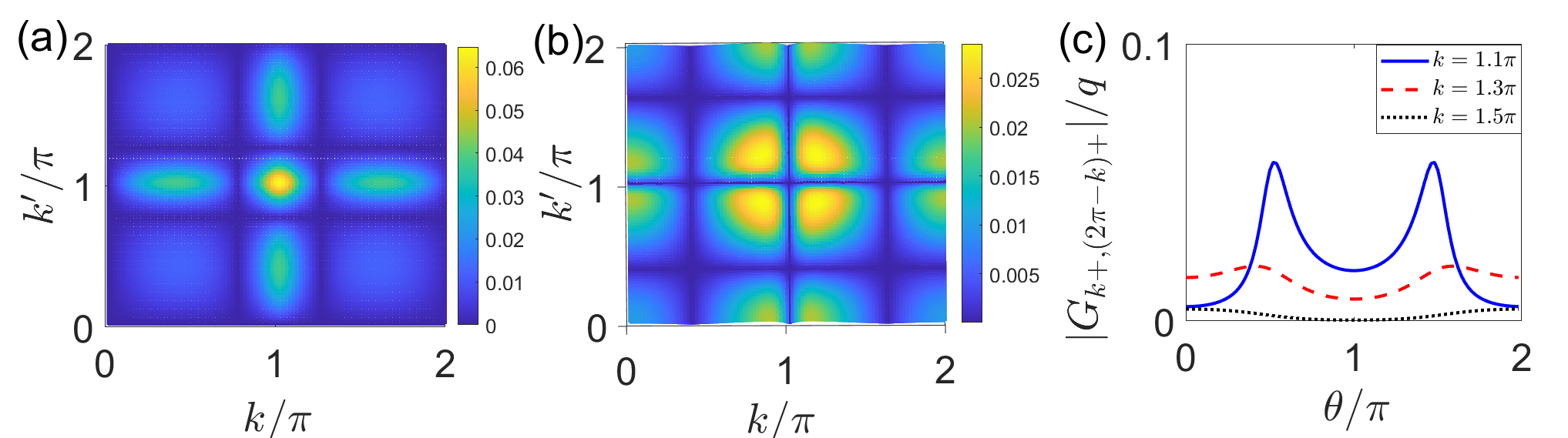}
  \caption{ $|G_{k-,k'-}|$ and $|G_{k+,k'+}|$ versus $k$ and $k'$ are plotted in (a) and (b) respectively for $\theta=0.4\pi$. (c) $|G_{k+,(2\pi-k)+}|$ versus $\theta$ for different $k$. The parameters for (a), (b) and (c) are $\theta=0.4\pi$, $L=100, n=50, m=51$ in the case of the A-B coupling.  The parameters shared by all of the figures are $\delta=0.5$ and $g=q$.}
\label{inband}
\end{figure*}

{For the parameters used in Fig.~~\ref{Energylevel}(d), a further numerical results show that the $(|G_{k+,k'+}|,|G_{k-,k'-}|)\gg(|G_{k+,k'-}|,|G_{k-,k'+}|)$ (not shown here) and therefore the inter-band coupling can be neglected. Moreover, $|G_{k-,k'-}|$ is nearly flat except near the regime of $(k,k')=\pi$ as shown in Fig.~\ref{inband}(a). Therefore, for the lower band of the SSH waveguide, the giant atom only trivially lifts the degeneracy [see Fig.~\ref{Energylevel}(d)]. On the contrary, at the regime of $\theta=0.4\pi$ where the degeneracy is lifted, we plot $|G_{k+,k'+}|$ in Fig.~\ref{inband} (b). It shows that the maximum value appears when $k+k'=2\pi$. We note that, for the bare SSH waveguide,
$E_{k+}=E_{(2\pi-k)+}$, therefore a strong effective coupling gives birth to the degeneracy lifted. A further evidence can be found in Fig.~\ref{inband}(c) where the dependence of $|G_{k+,(2\pi-k)+}|$ on $\theta$ is plotted for different $k$. It shows that for $k=1.1\pi$, the high peaks of $|G_{k+,(2\pi-k)+}|$ appear at $\theta\approx 0.4\pi$ and $\theta\approx 1.6\pi$, where the degenercy broken is always lifted as shown in Fig.~\ref{Energylevel}(d).  For $k=1.3\pi$ and $k=1.5\pi$, the value of $|G_{k+,(2\pi-k)+}|$ is relatively small compared to $k=1.1\pi$, therefore the degeneracy is not lifted when the energy level is far away from the lower boundary, which agrees with the energy spectra in Fig.~\ref{Energylevel}(d).}

\section{Topology of the System and Winding Number}
\label{Topology of the system and winding number}
The A-B coupling setup also allows us to study the topological property for $d=0$, in which the two-level giant atom is coupled to the waveguide through two different resonators in the same unit cell.

Different from the periodical boundary condition, the giant atom together with coupled two resonators forms an effective boundary.
{That is, the giant atom divides the SSH waveguide with periodical boundary condition into two waveguides with open boundary condition. However, when $d=0$, the length of one of which is zero and so we will get a single open SSH chain.}
In this case, we use the approach of the generalized Brillouin zone to study the non-Bloch winding number, {which can be calculated directly for the open SSH chain}~\cite{YSY}.
To this end, we take an ansatz of $(\phi_{n+l,A},\phi_{n+l,B})=\beta^l(\phi_{A},\phi_{B})\,(l>0)$,
with $\phi_{n+l,A}$ and $\phi_{n+l,B}$ being the single-photon excitation probabilities for the A and B resonators in the $(n+l)$th unit cell.
$\beta$ is a parameter to be determined.

{Considering the Hamiltonian
\begin{equation}
H=H_{\rm{SSH}}+\Omega|e\rangle\langle e|+g\left[\sigma_+(C_{A,n}+C_{B,n})+\rm{H.c.}\right],
\end{equation}
we have
 \begin{eqnarray}
(t_1+t_2\beta^{-1})\phi_B &=& E\phi_A,\nonumber \\
(t_1+t_2\beta )\phi_A&=&E\phi_B,
\label{eigen eqs}
\end{eqnarray}
via the Schr\"{o}dinger equation, where $l\in(1,L-1)$. Then we have
 \begin{eqnarray}
(t_1+t_2\beta^{-1})(t_1+t_2\beta)=E^2.
\label{eigen eqs2}
\end{eqnarray}
There are two solutions for $\beta$ with
 \begin{eqnarray}
 \beta_{\pm}(E)=\frac{E^2-t_1^2-t_2^2\pm\sqrt{(E^2-t_1^2-t_2^2)^2-4t_1^2t_2^2}}{2t_1t_2}. \nonumber\\
 \label{betapm}
\end{eqnarray}
 The corresponding eigenfunctions are
{\begin{eqnarray}
\phi_A^{(\pm)}=\frac{E}{t_1+t_2\beta_{\pm}}\phi_B^{(\pm)}
\label{phiA}
\end{eqnarray}}
or
{\begin{eqnarray}
\phi_B^{(\pm)}=\frac{E}{t_1+t_2\beta_{\pm}^{-1}}\phi_A^{(\pm)}.
\label{phiB}
\end{eqnarray}}}

\begin{figure}
  \centering
  \includegraphics[width=1 \columnwidth]{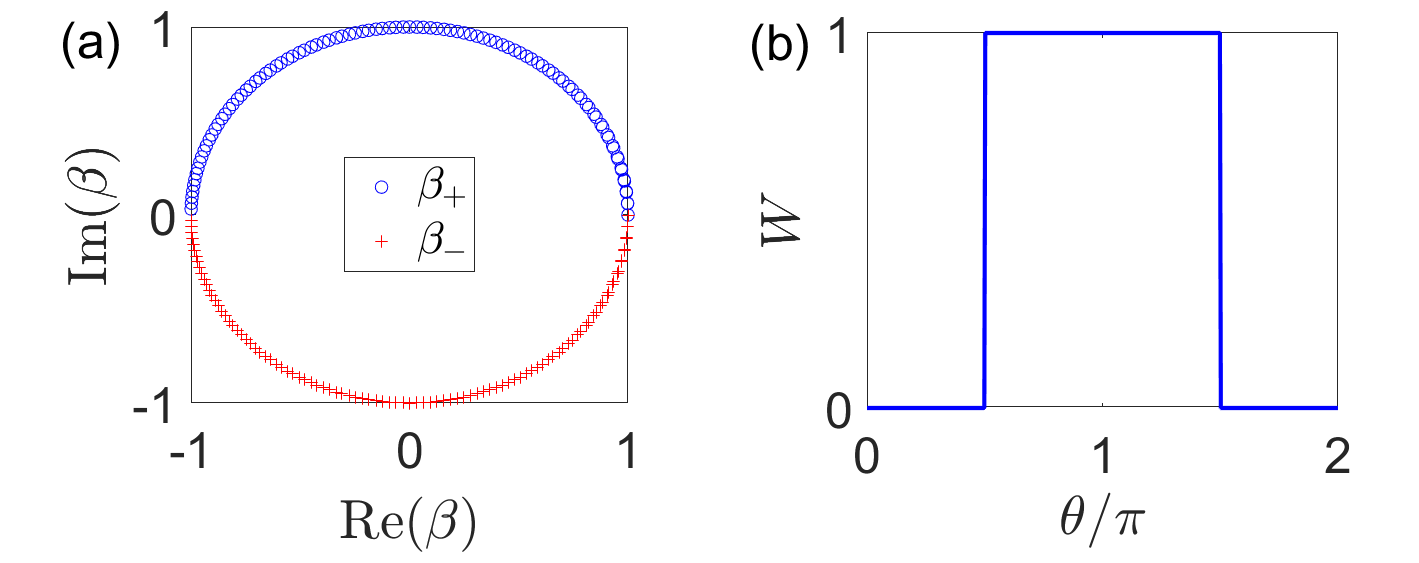}\nonumber\\
  \caption{(a) The real (imaginary) part of $\beta$ which includes $\beta_{-}$ and $\beta_{+}$ two branches. (b) The winding number versus $\theta$.
  The parameters are set as  $\delta=0.5q$, $\theta=0.8\pi$, $L=100, g=q, \omega=\Omega$ and $n=m=50$.}
\label{subbetawinding}
\end{figure}

{Then, the general solution of the wave function can be written as a linear combination
{ \begin{eqnarray}
\left(\begin{array}{c}
   \phi_{n+l,A} \\
   \phi_{n+l,B}
 \end{array}\right)=
 \beta_+^{l}\left(\begin{array}{c}
   \phi_{A}^{(+)} \\
   \phi_{B}^{(+)}
 \end{array}\right)+
  \beta_-^{l}\left(\begin{array}{c}
   \phi_{A}^{(-)} \\
   \phi_{B}^{(-)}
 \end{array}\right).
\label{linear combination}
\end{eqnarray}}
Substituting Eq.~\eqref{linear combination} into  the boundary condition, we obtain
\begin{widetext}
\begin{eqnarray}
&&t_1\left(\beta_+^{L-1}\phi_A^{(+)}+\beta_-^{L-1}\phi_A^{(-)}\right)+t_2\phi_x
=E\left(\beta_+^{L-1}\phi_B^{(+)}+\beta_-^{L-1}\phi_B^{(-)}\right),\nonumber\\
&&t_2\left(\beta_+^{L-1}\phi_B^{(+)}+\beta_-^{L-1}\phi_B^{(-)}\right)+t_1\phi_y+g\phi_c=E\phi_x\nonumber\\
&&t_1\phi_x+t_2\left(\beta_+\phi_A^{(+)}+\beta_-\phi_B^{(-)}\right)+g\phi_c=E\phi_y\nonumber\\
&&t_2\phi_y+t_1\left(\beta_+\phi_B^{(+)}+\beta_-\phi_B^{(-)}\right)
=E\left(\beta_+\phi_A^{(+)}+\beta_-\phi_A^{(-)}\right),\nonumber\\
&&g\left(\phi_x+\phi_y\right)=E\phi_c,
\label{boundary condition}
\end{eqnarray}
\end{widetext}
where $(\phi_x,\phi_y)=(\phi_{n,A},\phi_{n,B})$ represents the photonic excitation at the site of the giant atom.  Together with Eqs.~\eqref{phiA} and~\eqref{phiB}, we find that $\beta_+$ and $\beta_-$ satisfy the relation {$\alpha(\beta_+)=\alpha(\beta_-)$} where $\alpha$ is a function of $\beta$, and is expressed is
\begin{widetext}
\begin{eqnarray}
\alpha(\beta)=\frac{\beta^L\left(\frac{g^2}{\sqrt{(t_1+t_2\beta)(t_1+t_2\beta^{-1})}}-t_1\right)
+\left(\frac{g^2}{t_1+t_2\beta}+t_1\right)}
{\beta^L\left(\frac{g^2}{\sqrt{(t_1+t_2\beta)(t_1+t_2\beta^{-1})}}+t_1\sqrt{\frac{t_1+t_2\beta^{-1}}{t_1+t_2\beta}}\right)
+\left(\frac{g^2}{t_1+t_2\beta}-t_1\sqrt{\frac{t_1+t_2\beta^{-1}}{t_1+t_2\beta}}\right)}.
 \label{beta12}
\end{eqnarray}
\end{widetext}}
{Actually, under the periodical boundary condition, $\beta_{\pm}$ is nothing but $\exp(\pm ik)$. For the bulk state inside the continuous band with the presence of the giant atom,  that is, $(t_1-t_2)^2<E^2<(t_1+t_2)^2$ in our model, we obtain {$|\beta_+|=|\beta_-|=1$}. In Fig.~\ref{subbetawinding}(a), we numerically plot {$\beta_{\pm}$} in the complex plane which forms the generalized Brillouin zone. Here, we emphasize that although the translation invariance is broken by the giant atom, the whole system is still Hermitian, and the generalized Brillouin zone also forms a unit circle, which is different from that for the non-Hermitian system~\cite{YSY}.}

{With the assistance of generalized Brillouin zone, we can calculate the winding number. We define the non-Bloch Hamiltonian
\begin{eqnarray}
H(\beta)=(\begin{array}{cc}
            0 & t_1+t_2\beta^{-1} \\
            t_1+t_2\beta & 0
          \end{array}
)
 \label{non-Bloch Hamiltonian}
\end{eqnarray}
which is obtained from Eq.~\eqref{eigen eqs}. The Hamiltonian is consistent with the SSH model when $\beta$ is replaced by $e^{ik}$. Following the procedure in Ref.~\cite{YSY},  we define the ``Q matrix":
\begin{eqnarray}
Q(\beta)=|\tilde{\mu}(\beta)\rangle\langle \tilde{\mu}(\beta)|-|\mu(\beta)\rangle\langle \mu(\beta)|
 \label{Q matrix}
\end{eqnarray}
where $|\mu(\beta)\rangle$ is the eigenstate of $H(\beta)$ and $|\tilde{\mu}(\beta)\rangle=\sigma_z|\mu(\beta)\rangle$. The $Q$ matrix is off-diagonal with the structure
\begin{equation}
Q=\left(\begin{array}{cc}
                0 & q \\
                q^{-1} & 0
              \end{array}\right).
\end{equation}}

 Integrating the $q$ function with $q=-\sqrt{(t_1+t_2\beta^{-1})/(t_1+t_2\beta)}$, we can apply the non-Bloch winding number~\cite{YSY}
\begin{eqnarray}
W=\frac{i}{2\pi}\int_{C_{\beta}}q^{-1}dq,
 \label{non-Bloch winding number}
\end{eqnarray}
 to demonstrate the topology property of the whole system, where the integral is performed on the loop $C_\beta$.  In Fig.~\ref{subbetawinding}(b), we plot $W$ as a function of $\theta$, and show the topological phase transition at $\theta=\pi/2$ and $\theta=3\pi/2$, which is similar to that of the bare SSH waveguide. In other words, the giant atom does not change the winding number of the bulk state. The A-A coupling or the A-B coupling with $d\neq0$ are very different from the traditional open boundary,  and therefore we cannot use the non-Bloch winding number to discuss the topology.
{ We emphasize this analogical properties between the traditional open boundary and the atom-type open boundary just shows the boundary effect of the giant atom.}


\section{Zero Mode and Bound States}
\label{Sec Zero mode and bound states}
\subsection{Analytical Solutions for Zero Modes}
\label{Analytical Solutions for Zero Modes}
\begin{figure}
\centering
\includegraphics[width=8cm]{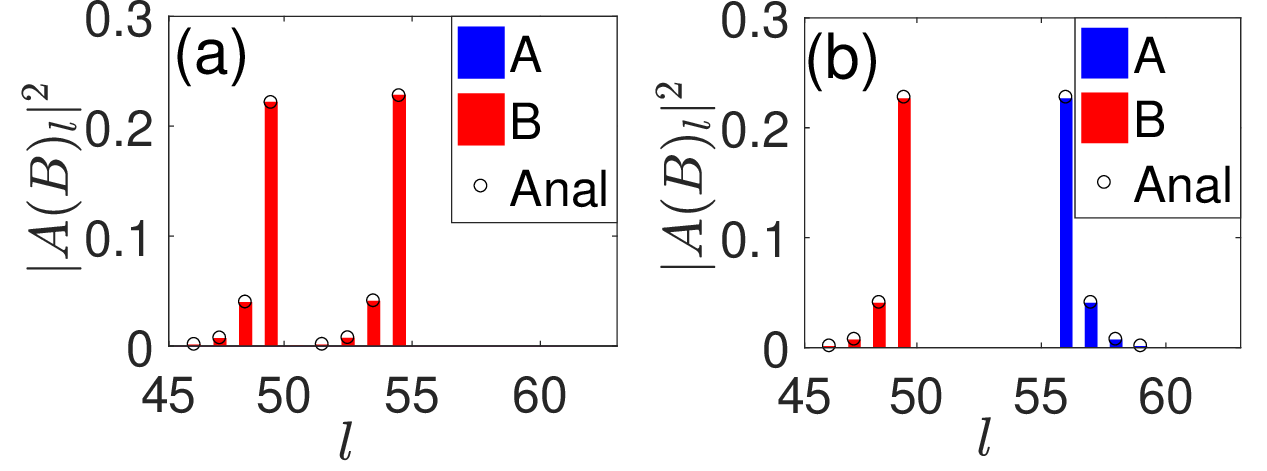}
\includegraphics[width=8cm]{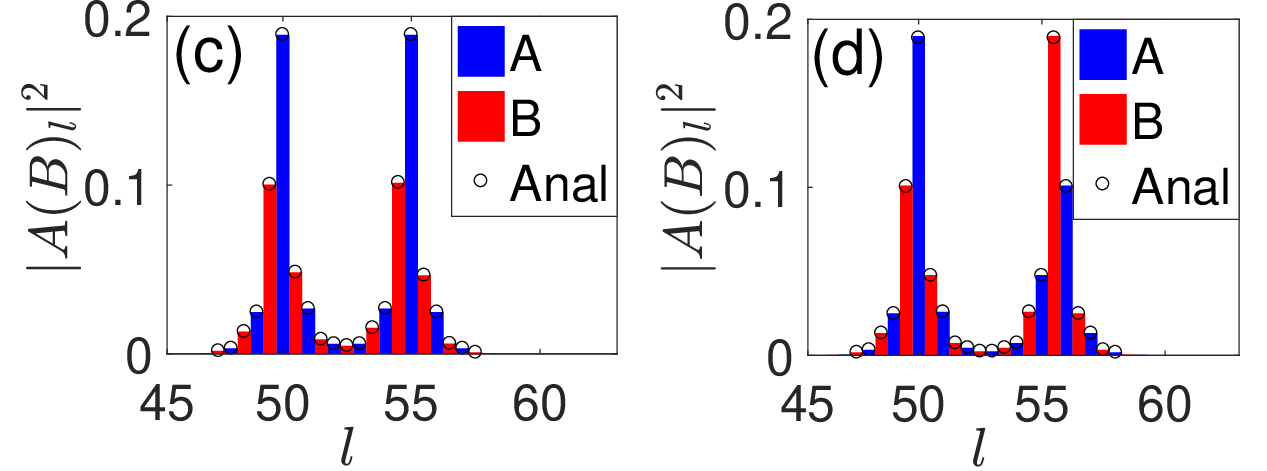}
\caption{ Photon distribution of zero mode in (a) for A-A and (b) for A-B coupling.  Photon distribution in the upper bound state in (c) for the A-A and (d) for the A-B coupling.
The bars are numerical results and the empty circles are the analytical ones. The parameters are chosen as $L=100, n=50, m=55$, $\theta=0.8\pi, \delta=0.5$ and $g=q$.}
\label{Zmode}
\end{figure}

In Fig.~\ref{Zmode}, we show the photonic spatial distributions for the zero energy and bound states, which are both the atom-waveguide dressed states, and bound states locate outside the upper and lower energy bands. Here, we focus on the topologically nontrivial phase {($t_2> t_1, \pi/2<\theta<3\pi/2$)} and the physics for the topologically trivial phase can be found in Appendix~\ref{Zero mode and bound states}. In Fig.~\ref{Zmode}, the bars and empty circles denote the numerical and the analytical results, respectively. The analytical solutions are obtained as follows.

For the zero mode corresponding to $E=0$ in the topologically non-trivial phase under the condition of the A-A coupling, the photon population can be obtained in Appendix~\ref{Zero mode and bound states} as $A_l=0$ and
 \begin{eqnarray}
\frac{B_{l}}{U_{e}}&=&Y_2\times
\begin{cases}
(-\tau)^{-(l-n)}+(-\tau)^{-(l-m)} & (l<n)\\
(-\tau)^{-(l-m)} & (n\leq l<m)\\
0 & (m\leq l)
\end{cases},\nonumber \\&& \label{eq:8}
\end{eqnarray}
where $A_l (B_l)$ denotes the photonic amplitude of probability in $A$-type ($B$-type) resonator of the $l$th unit cell, and we have defined $Y_2=g/t_{1}$ and $\tau=t_1/t_2$. Equation~(\ref{eq:8}) shows that the photons only occupy $B$-type resonators,  whose label numbers are smaller than $m$ {and satisfies the spatial symmetry $|B_r|=|B_{r+m-n}|$ with $r<n$}. That is, as shown in Fig.~\ref{Zmode}(a), the photons only distribute in the resonators in the left side of the giant atom or between two coupling points. However, for the A-B coupling, the eigen state of the zero mode can be obtained as (Appendix~\ref{Zero mode and bound states})
\begin{eqnarray}
\frac{A_l}{U_e}&=&Y_2\times\begin{cases}(-\tau)^{(l-m)}&(l> m)\\
0& (l\leq m)\end{cases},\nonumber \\
\frac{B_l}{U_e}&=&
Y_2\times\begin{cases}(-\tau)^{-(l-n)}& (l< n)\\
0& (l\geq n)\end{cases},
 \label{SAB1}
\end{eqnarray}
which shows that the photons occupy \textit{B}(\textit{A})-type resonators on the left (right) side of the giant atom. The photon distributions satisfy the spatial symmetry $|A_r|=|B_{m+n-r}|$, but the resonators between two coupling points are not occupied.  This has been clearly shown in Fig.~\ref{Zmode}(b). That is, the photon occupations for zero mode have chirality for both the A-A and A-B couplings.

For the two bound states, the photon distributions are lack of the chirality, as shown in Figs.~\ref{Zmode}(c) and (d), where the photon distributions for the upper bound state are plotted for the A-A and A-B couplings, respectively. Figs.~\ref{Zmode}(c) and (d) show that the photons occupy both $A$- and $B$-type resonators in the chain, and the occupation probabilities decay exponentially around the two coupling points. More interestingly, as shown in Figs.~\ref{Zmode}(c) and (d), the bound states satisfy the spatial symmetry $|A_r|=|A_{m+n-r}|, |B_r|=|B_{m+n-r}|$ for the A-A coupling and $|A_r|=|B_{m+n-r}|,|B_r|=|A_{m+n-r}|$ for the A-B coupling. Such symmetry can also be analytically verified as shown in Appendix~\ref{Zero mode and bound states}.

\subsection{Probing Zero Modes}

In the Sec.~\ref{Energy spectrum equations}, we have demonstrated that the coupling to the giant atom modifies the energy structure of the SSH waveguide. In principle, the energy structure can be detected by the scattering approach, that is, when the incident particle is resonant with the system, it is completely reflected~\cite{Landau}. Another issue is how to detect the profile of the wave function. Here, we propose two approaches to deal with such problem.

\begin{figure}
  \centering
  \includegraphics[width=8cm]{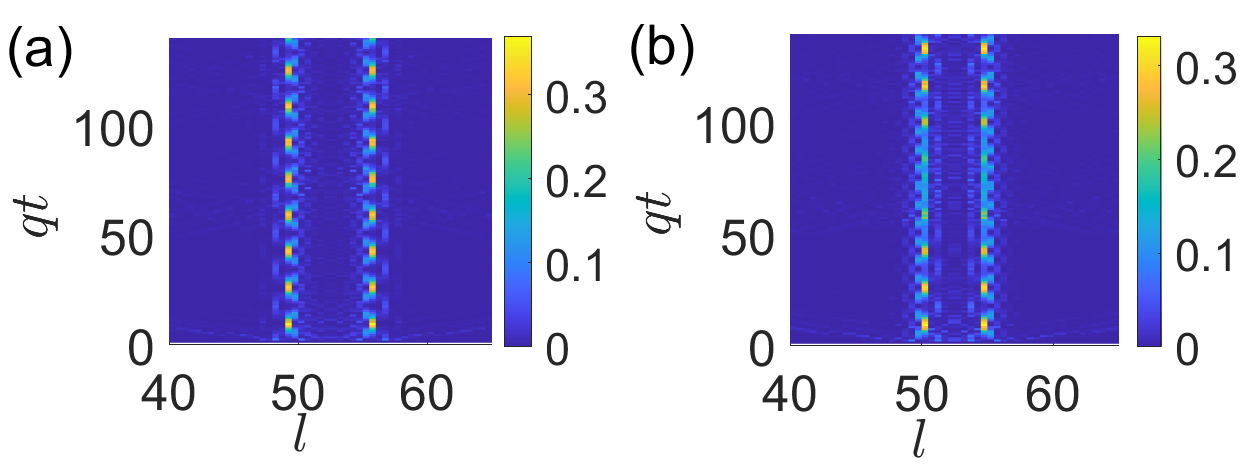}\nonumber\\
  \includegraphics[width=8cm]{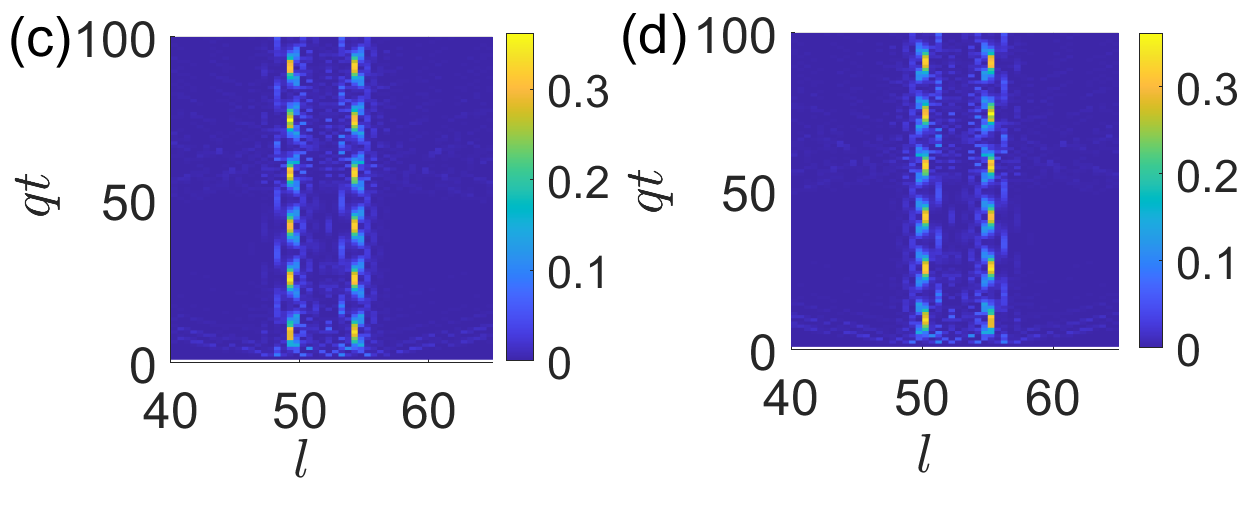}\nonumber\\
  \caption{The evolution of the photon in the SSH chain coupling with the giant atom via the A-B coupling in (a) and (b) or the A-A coupling in (c) and (d).
  The parameters are set as $(\Omega-\omega)/q=0, \delta=0.5q, L=100, g=q, n=50$ and $m=55$. $\theta=0.8\pi$ for (a), (c) and $\theta=0.2\pi$ for (b), (d).}
\label{qw}
\end{figure}

In the first approach, we prepare the giant atom in its excited state initially, and observe the photonic distributions during the emission of the atom by setting $\omega=\Omega$. Remember that the zero mode exists in the A-A coupling setup and topologically nontrivial phase for the A-B coupling setup. In these situations, we have exhibited the evolution of the photonic state in the waveguide in Figs.~\ref{qw}(a), (c) and (d). They show that the photons are bounded in the coupling points and behave as an oscillation function (see the sites of $l=50$ and $l=55$) and the oscillation frequency is determined by the atom-waveguide coupling  strength $g$. A more detailed observation also permits the oscillation in the adjacent sites, which manifests the chiral nature of the zero mode. For example, in Fig.~\ref{qw}(a), we find the oscillation in the sites outside the giant atom [$l=49$ and $l=56$ ], and on the left (right) and inside the giant atom in Figs.~\ref{qw} (c) ($l=49$ and $l=54$) and (d) ($l=51$ and $l=56$),  respectively. These results predict the profile of the zero mode, which is given in Figs.~\ref{boundstate13} and ~\ref{boundstate24} in Appendix.~\ref{Zero mode and bound states} and Figs.~\ref{Zmode}(a) and (b) in Sec.~\ref{Analytical Solutions for Zero Modes} of the main text. For the A-B coupling, in the topologically trivial phase, the eigenstate located in the gap detunes with the giant atom, and the photonic distribution is demonstrated in Fig.~\ref{qw}(b). The detuning is manifested by the two distinguishable frequencies.

As for the second approach, we introduce an auxiliary probing atom, which is resonant with the zero mode, and observe its excitation evolution. Taking the A-A coupling as an example, we can write the Hamiltonian as
\begin{equation}
H=H_{\rm{SSH}}+H_{I,{\rm AA}}+f(\tau_+C_{B,l}+{\rm H.c.})-i\gamma|e\rangle_p\langle e|
\end{equation}
which demonstrates that the probing atom is coupled to the site B in the $l$th unit cell with the coupling strength $f$. $\gamma$ and $\tau_+=|e\rangle_p\langle g|$ are the spontaneous emission and raising operator of the probing atom.

\begin{figure}
  \centering
  \includegraphics[width=8cm]{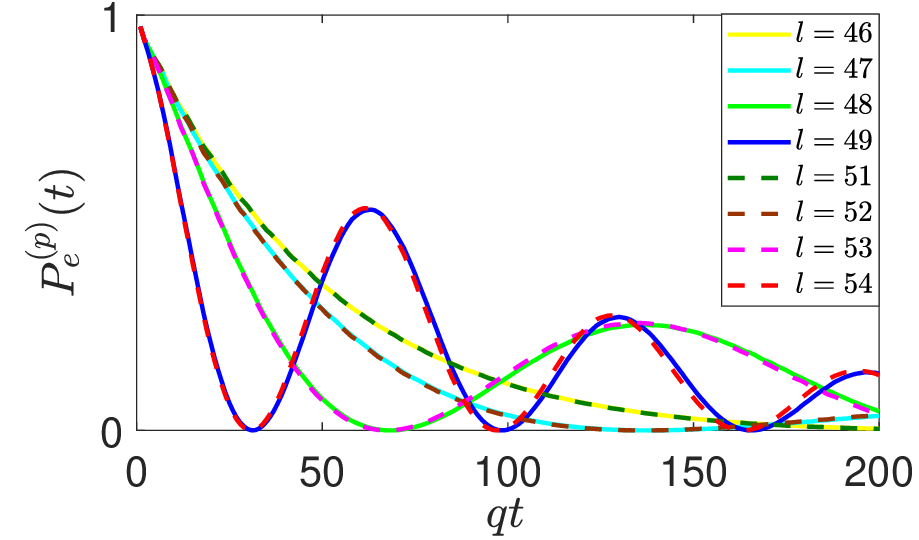}
  \caption{The evolution of the population in the excited state for the probing atom. The parameters are set as {$L=100, n=50, m=55, \delta=0.5, g=q$,$f=0.1q$ and $\gamma=0.2q$}. }
\label{probe}
\end{figure}

Preparing the probe atom in the excited state, while the giant atom and the waveguide in their ground states initially, we plot the evolution of the excited state population $P_e^{(p)}(t)=\langle|e\rangle\langle e|\rangle_p$ of the probe atom in Fig.~\ref{probe} when it is coupled to different sublattices. As shown in Fig.~\ref{probe}, the population shows an obvious Rabi oscillation when the probe atom is coupled to the site near the atom-waveguide connecting points. On the contrary, the population undergoes an exponential decay, which is dominated by the spontaneous emission of the probe atom when it locates far away from the connecting points.

The underlying physics can be explained by a single-mode light-matter interaction model. Since the probe atom is largely detuned from the bulk energy bands, it only exchanges energy with the zero mode of the probed system, and the effective Hamiltonian can be expressed as
\begin{equation}
H=-i\gamma|e\rangle_p\langle e|+(\xi_l\tau_+|z\rangle\langle G|+{\rm H.c.} )
\end{equation}
where $|z\rangle$ is the zero mode and {$\xi_l=f\langle G|C_{B,l}|z\rangle=fB_l$}. As a result, the period of the Rabi oscillation in Fig.~\ref{probe} is $T_l=2\pi/|\xi_l|$.  In this sense, the dynamical behavior of $P_e^{(p)}(t)$ directly witnesses that the photon distributes nearby the coupling points for the zero mode.
Furthermore, the coincidence of some curves (for example $l=48$ and $l=53$) manifests the symmetry given in Sec.~\ref{Analytical Solutions for Zero Modes} [also see Eq.~(\ref{SAB1})].

\subsection{Robustness of the Zero Mode}
We further study the robustness of the zero mode and bound states to the disorder, atomic dissipation, the asymmetric atom-waveguide coupling as well as the next nearest neighbour coupling in the SSH waveguide (Appendix~\ref{Robustness of zero modes}). We find that the topological protection makes the chirality of zero mode more robust to the disorder and asymmetric atom-waveguide coupling, but the symmetries of the bound states are lost. Furthermore, considering the atomic dissipation in a non-Hermitian manner, the real part of the energy spectrum can also reveal the topological nature of the system. However, by introducing the next nearest neighbour coupling, the zero mode does not exist anymore even that the two-level giant atom is coupled to the topological waveguide, the energy spectrum in this situation is similar to that of the generalized SSH model with the next nearest neighbour coupling and open boundary~\cite{LL}.

\section{Energy Band}

{In this section, we discuss the states inside the energy band. We consider an energy level which localizes inside the energy band. Such energy level can be mathematically characterized by  $E^2=t_1^2+t_2^2+2t_1t_2\cos(\lambda)$, where $\lambda$ is an arbitrary real number. Then we have
\begin{eqnarray}
f(k)=\frac{1}{2[\cos(\lambda)-\cos(k)]}
\approx \frac{1}{2\lambda}\left(\frac{1}{k-\lambda}-\frac{1}{k+\lambda}\right),\nonumber\\
 \label{fk2}
\end{eqnarray}
according to the definition of $f(k)$ in Eq..~(\ref{fk}), where the approximate condition holds when $\lambda,k\rightarrow 0$. For the A-A coupling, according to Eqs.~(\ref{AAstateL}) and ~(\ref{fk2}), we have
{\begin{eqnarray}
\frac{A_l}{U_e}\approx\frac{-T}{\lambda}\cos\left[\frac{\lambda(m-n)}{2}\right]\sin\left[\lambda\left(l-\frac{n+m}{2}\right)\right]\nonumber\\
 \label{bandstateA}
\end{eqnarray}}
and
{\begin{eqnarray}
&\frac{B_l}{U_e}\approx -\cos\left[\frac{\lambda(m-n)}{2}\right]\times&\nonumber\\
&\left\{\frac{Y_1}{\lambda}\sin\left[\lambda\left(l-\frac{n+m}{2}\right)\right]
+\frac{Y_2}{\lambda}\sin\left[\lambda\left(l-\frac{n+m+2}{2}\right)\right]\right\},&\nonumber \\
 \label{bandstateB}
\end{eqnarray}}}
where $T=gE/(t_1t_2)$ and $Y_1=g/t_{2}$.

\begin{figure}
  \centering
  \includegraphics[width=8cm]{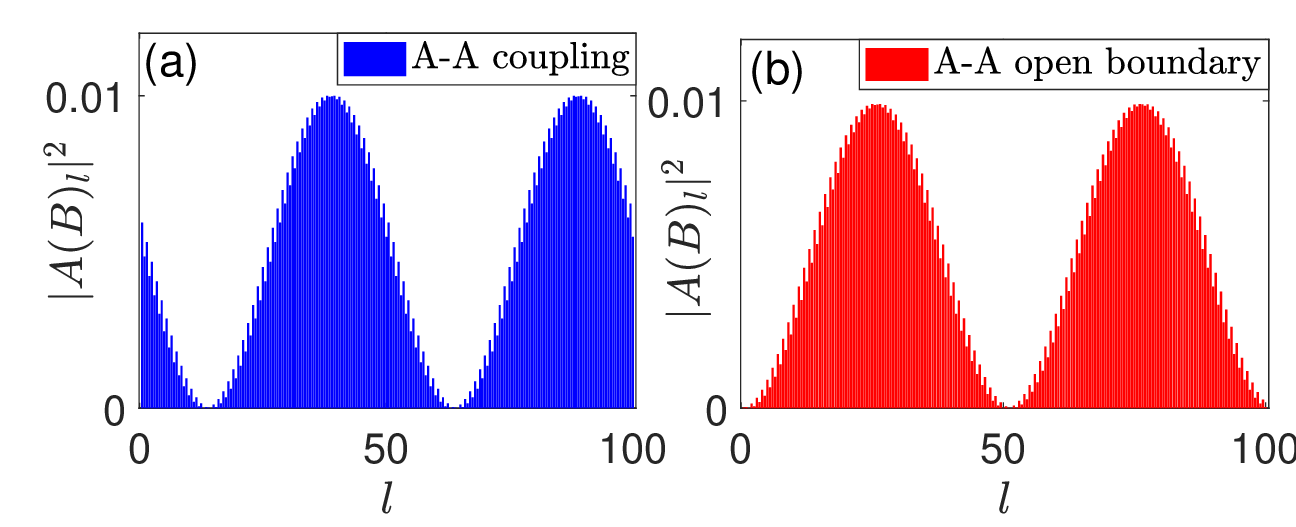}\nonumber\\
    \includegraphics[width=8cm]{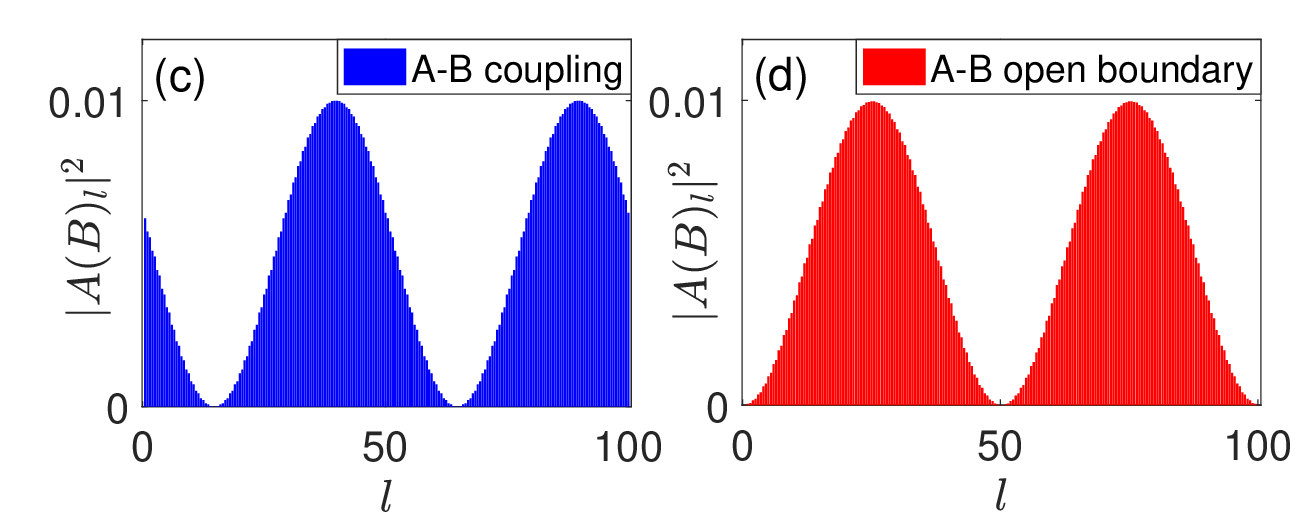}
  \caption{The photon distribution for the second top band energy level under the A-A and A-B coupling, respectively in (a) and (c). The photon distribution for the second top band energy level under the A-A and A-B open boundary setup, respectively in (b) and (d).
  The parameters are set as  $L=101$ for (b) and  $L=100$ for (a), (c) or (d),  $n=50, m=80$, $\theta=0.8\pi$, $\delta=0.5$  and $g=q$. }
\label{bandstatex}
\end{figure}

{The above results show that the photon distributes along the waveguide with a sinusoidal shape, as shown in Fig.~\ref{bandstatex}(a) for the top energy level in the upper band. As a comparison, we also plot the photon distribution for the open SSH waveguide with odd sites in Fig.~\ref{bandstatex}(b). The similarity between these two figures also shows the effective boundary effect of the giant atom.}

{It is obvious that, at the site $l=(m+n)/2$, the excitation amplitude is zero for the A-A coupling. However, this is not the case for the A-B coupling. For the latter one, we plot the photon distribution in Fig.~\ref{bandstatex}(c), whose approximate result is obtained as
{\begin{eqnarray}
\frac{A_l}{U_e}
&\approx&-\frac{T}{2\lambda}\sin[\lambda(l-n)]\nonumber \\
&&-\left\{\frac{Y_1}{2\lambda}\sin[\lambda(l-m)]
+\frac{Y_2}{2\lambda}\sin[\lambda(l-m-1)]\right\},\nonumber \\
 \label{bandstateA1}
\end{eqnarray}}
\begin{eqnarray}
\frac{B_l}{U_e}
&\approx&-\frac{T}{2\lambda}\sin[\lambda(l-m)]\nonumber \\
&&-\left\{\frac{Y_1}{2\lambda}\sin[\lambda(l-n)]
+\frac{Y_2}{2\lambda}\sin[\lambda(l-n+1)]\right\}.\nonumber \\
 \label{bandstateB1}
\end{eqnarray}
The result for a bare open SSH waveguide with even sites is given in Fig.~\ref{bandstatex}(d). Again, the boundary effect of the giant atom is demonstrated clearly. At last, we emphasize that Eq.~\eqref{fk2} becomes unreasonable when the band level $E$ are between two adjacent energy levels $\omega_k$s, and also Eqs. ~\eqref{bandstateA}, ~\eqref{bandstateB}, ~\eqref{bandstateA1} and therefore ~\eqref{bandstateB1} are no longer applicable.}


\section{Non-Markovian Retardation Effect}
\label{Non-Markovian retardation effect}
\begin{figure}
  \centering
  \includegraphics[width=8cm]{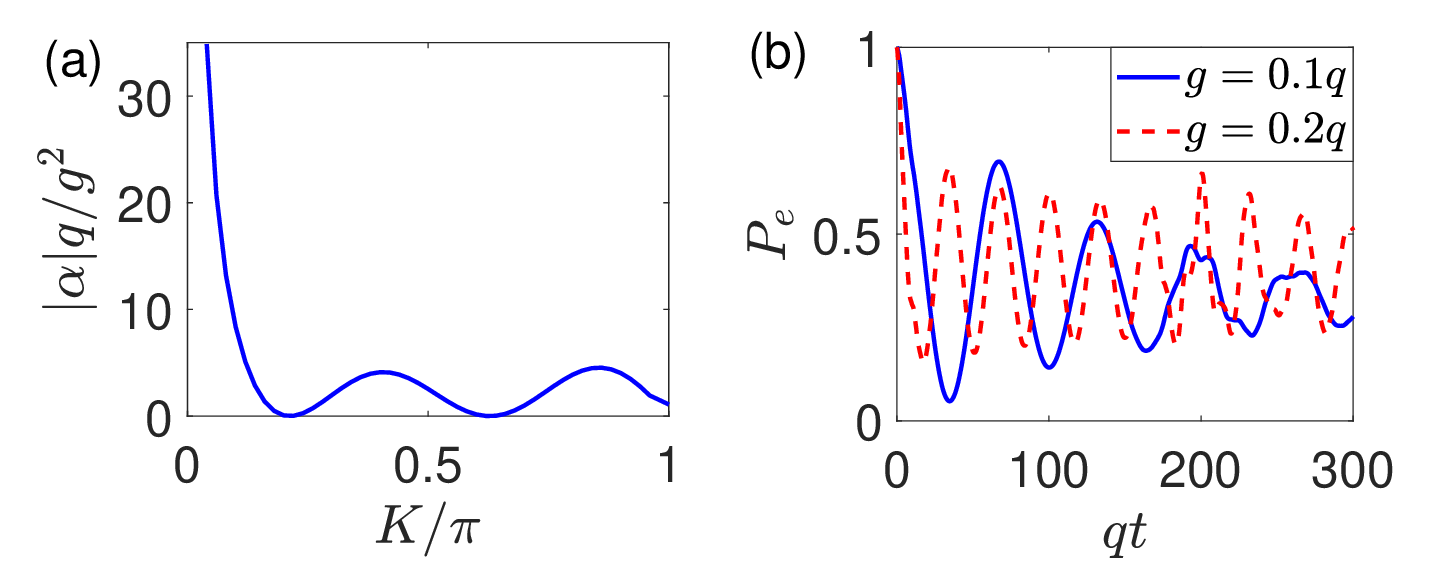}\nonumber\\
    \includegraphics[width=8cm]{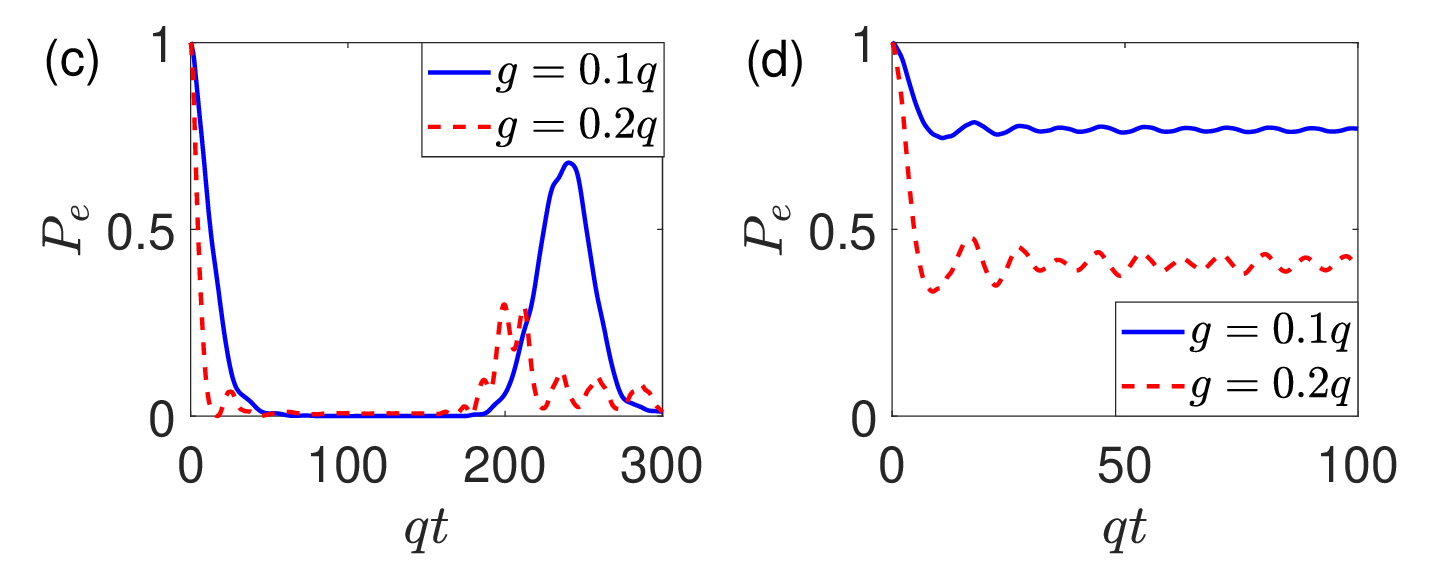}\nonumber\\
  \caption{(a) The decay rate $|\alpha|$ as a function of $K$. (b), (c) and (d) The evolution of atomic population $P_e$ for the A-B coupling setup. The parameters are set as $\delta=0.5q, L=100$, $n=50$, $m=54$ and $\theta=0.8\pi$. $(\Omega-\omega)/q=1.99\, (K \approx 0.06\pi)$ for (b).  $(\Omega-\omega)/q=1.66\, (K\approx 0.4\pi)$ for (c). $(\Omega-\omega)/q=1.3\, (K\approx 0.62\pi)$ for (d).}
\label{markovian2}
\end{figure}
Besides the chirality of the zero mode, the topology of the system can also be detected by the dynamical evolution of the giant atom. If we consider the SSH waveguide as a structured environment, then the giant atom is not affected by the waveguide when its frequency is in the energy gap of the waveguide. However, when the frequency of the giant atom is inside the energy band, the dynamics of the giant atom obeys the master equation (for the detailed derivations, please refer to Appendix~\ref{Emission of giant atom}).
\begin{equation}
\frac{d\rho}{dt}=(\alpha+\alpha^{*})\sigma^{-}\rho\sigma^{+}-\alpha \rho\sigma^{+}\sigma^{-}-\alpha^{*}\sigma^{+}\sigma^{-}\rho,
\label{markovian}
\end{equation}
with the Markovian approximation. For the A-A coupling, $\alpha=g^2[1+\cos K(n-m)]/|v_g(K)|$ with $\Omega-(\omega\pm\omega_K)=0$ for upper ($+$) and lower ($-$) band, respectively. However, for the A-B coupling, we obtain $\alpha=g^2(\omega_K\pm Q(K))/(|v_g(K)|\omega_K)$ with $\Omega-(\omega\pm\omega_K)=0$  for upper ($+$) and lower ($-$) band, respectively. Here, $Q(k)$ is defined below Eq.~(\ref{eq:5}) and $v_g(k)=\partial \omega_k/\partial k$ represents the group velocity of the travelling photon in the waveguide with wave number $k$. We here only consider the zero temperature situation, and the discussion of the temperature effect can be found in Appendix~\ref{Temperature effect}. For the A-B coupling setup, we plot $|\alpha|$, which demonstrates the decay rate of the giant atom, as a function of $K$ in Fig.~\ref{markovian2}(a). It shows that the decay rate will acquire a
large value on the edge of the band ($K\approx 0$) and otherwise very small or even achieves zero for certain $K$. Governed by the Markovian master equation in Eq.~(\ref{markovian}), the atomic population $P_e=\langle |e\rangle \langle e|\rangle$ undergoes an exponential decay. To investigate the non-Markovian effct, we here plot the exact atomic dynamics based on the Sch\"{o}dinger equation in Figs.~\ref{markovian2}(b), (c) and (d), in which the atom is excited while all of the resonators in the waveguide are in the vacuum state initially, and the divergence from the exponential decay reveals the non-Markovian effects.

The non-Markovian effect here comes from the following two aspects. On the one hand, as the frequency of the atom is nearby the edge of energy band, the giant atom effectively couples to a single collective mode in the waveguide (Appendix~\ref{Non-Markovian Retardation effect}),  and the dynamics is characterized by the Rabi oscillation, whose frequency is proportional to the atom-waveguide coupling strength $g$ as shown in Fig.~\ref{markovian2}(b). In this situation, the non-Markovian effect comes from the ``frozen'' of the emitted photon, whose group velocity is nearly zero. This kind of non-Markovian effect is common for a waveguide system with band structure~\cite{photon1,photon2}. On the other hand, when the giant atom is resonant with the level inside the energy band, the finite group velocity of the emitted photon will induce the retardation effect. In Fig.~\ref{markovian2}(c), we consider $(\Omega-\omega)/q=1.66\, (K\approx 0.4\pi)$, which corresponds a relatively large $|\alpha|$. It shows that the emitted photon by the giant atom can excite the atom after they travels along the whole waveguide and come back to the atomic site, this kind of the retardation effect is common when a giant or small atom interact with a waveguide with periodical boundary condition. Here, a stronger atom-waveguide coupling induces minor oscillations. It implies that the atom effectively interacts with a few photonic modes with different group velocities due to the relatively strong atom-waveguide coupling.
Furthermore, for the situation of $\alpha\approx 0$ with $(\Omega-\omega)/q=1.3\, (K\approx 0.62\pi)$,  the emitted photons from one atom-waveguide coupling point will arrive the other point and induce the atomic oscillation as shown in Fig.~\ref{markovian2} (d), in which the oscillation frequency is nearly independent of $g$. This kind of retardation effect is only for the giant atom setup and we can observe it clearly when $|n-m|/v_g(K)\ll 1/|\alpha|$. In Appendix~\ref{Non-Markovian Retardation effect}, we further demonstrate the average photon numbers in each resonator and the photonic dynamics coincides with the above physical processes for different $\Omega$.


\section{Discussion and Conclusions}
\label{Discussion and conclusions}
We have studied the interaction between a giant atom and a topological waveguide, which is described by the SSH chain formed by the resonators with periodical boundary condition. The giant atom can act as an effective boundary to induce a chiral zero mode when it is coupled to two same-type resonators in the SSH chain, and lifts two-fold degeneracy of all bulk states in all parameter regimes of the SSH waveguide. However, if the giant atom is coupled to two different types of the resonators in the SSH chain, we find that the zero mode appears in topologically nontrivial regime and there is a nonzero mode inside the energy gap in the non-topological regime.{These features suggest that the giant atom acts the effective boundary and this atom-type boundary exhibits unique performance in the SSH chain.}
In addition, as a structured environment, the waveguide induces the giant atom to have the non-Markovian dynamical evolution and the retardation effect.

Our study can be applied to any system consisting a giant atom or an impurity coupled to the topological waveguide via two coupling points. We now discuss one of possible experimental realizations using superconducting quantum circuits, in which the superconducting qubit and the LC resonator chain act as giant atom and SSH chain, respectively. In such system, the coupling strength between resonators or between resonator and giant atom can be $50\sim 200$ MHz for the current technology~\cite{SHV,PRet,RMBS}. Moreover, the coupling between different superconducting elements can be experimentally tuned using various methods, e.g., periodical modulations~\cite{TGK}. That is, the change from trivial to non-trivial topological interaction can be experimentally observed by tuning the coupling between resonators.
The chiral and symmetry nature of the zero mode can be detected by introducing an auxiliary probing atom.  For the coupling strength $100$ MHz between resonators or between the resonators and giant atom, the Rabi oscillation can be observed for the zero mode when the coupling strength between the probe atom and the zero mode is $10$ MHz within the decay time $T_1=10\,{\rm \mu s}$ of the giant atom~\cite{MKM} for even larger decay rate of the probing atom.

In summary, we study the light-matter interaction in the context of the topological waveguide QED. Our study is also possible to be experimentally realized in photonic and other solid state systems. The analytical treatment for one-dimensional system lays a solid basis to understand the quantum effects in high-dimension topological waveguide systems and is helpful to further study the interaction between matter and topological environment.

Z. W. is supported by National Key R$\&$D Program of China (No. 2021YFE0193500), the National Natural Science Foundation of China (Grant Nos. 11875011 and 12047566); Y. L. is supported by NSFC under Grant No. 11874037, and Key-Area R\& D Program of GuangDong Province under Grant No. 2018B030326001.

\appendix
\section*{}

\section{Su-Schrieffer-Heeger Model}
\label{Su-Schrieffer-Heeger model}

We begin with a 1D waveguide which is described by Su-Schrieffer-Heeger (SSH) model, the Hamiltonian is written  as~\cite{WP}
\begin{equation}\label{Hsshl}
H_{\rm{SSH}}=\sum_{l}\left[t_1\hat{C}^{\dag}_{A,l}\hat{C}_{B,l}+t_2\hat{C}^{\dag}_{A,l+1}\hat{C}_{B,l}\right]+\rm{H.c.},
\end{equation}
where $\hat{C}_{A(B),l}$ is annihilation operator of the site A (B) in the $l$th unit cell. The coupling strengths are {$t_1=q(1+\delta \cos\theta)$ and $t_2=q(1-\delta \cos\theta)$ }with $\delta$ being the dimerization strength and $\theta$  can vary from 0 to $2\pi$ continuously.
By applying the Fourier transform under the periodic boundary condition
\begin{equation}
\hat{C}_{\alpha,l}=\frac{1}{\sqrt{L}}\sum_k e^{ikl}\hat{C}_{\alpha,k},
 \label{Fourier transformation}
\end{equation}
to Eq.~(\ref{Hsshl}) with $\alpha=A,\,B$, we have
\begin{equation}
H_{\rm{SSH}}(k)=\sum_{k}\psi(k)^{\dagger}h(k)\psi(k),
 \label{Hsshl(k)}
\end{equation}
where
\begin{equation}
\psi(k)=\left(
          \begin{array}{c}
            \hat{C}_{A,k} \\
            \hat{C}_{B,k}\\
          \end{array}
        \right),
\end{equation}
\begin{equation}
        h(k)=\left(
        \begin{array}{cc}
          0 & t_1+t_2\exp(-ik) \\
          t_1+t_2\exp(ik) & 0 \\
        \end{array}
      \right).
\label{h(k)}
\end{equation}
Then, the energy spectrum can be given as
\begin{equation}
E_{k\pm}=\pm \sqrt{t_1^2+t_2^2+2t_1t_2\cos(k)}=\pm\omega_k,
 \label{energy spectrum}
\end{equation}
and the corresponding eigen-states are
\begin{equation}
\left|E_{k\pm}\right\rangle=
\left(\frac{t_1+t_2e^{-ik}}{\sqrt{2}\omega_k}{\hat C}_{A,k}^{\dag}\pm\frac{1}{\sqrt{2}}{\hat C}_{B,k}^{\dag}\right)
\left|G\right\rangle,
 \label{statek0}
\end{equation}
where $\left|G\right\rangle$ is the ground state of the waveguide.

\begin{figure}
  \centering
  \includegraphics[width=8cm]{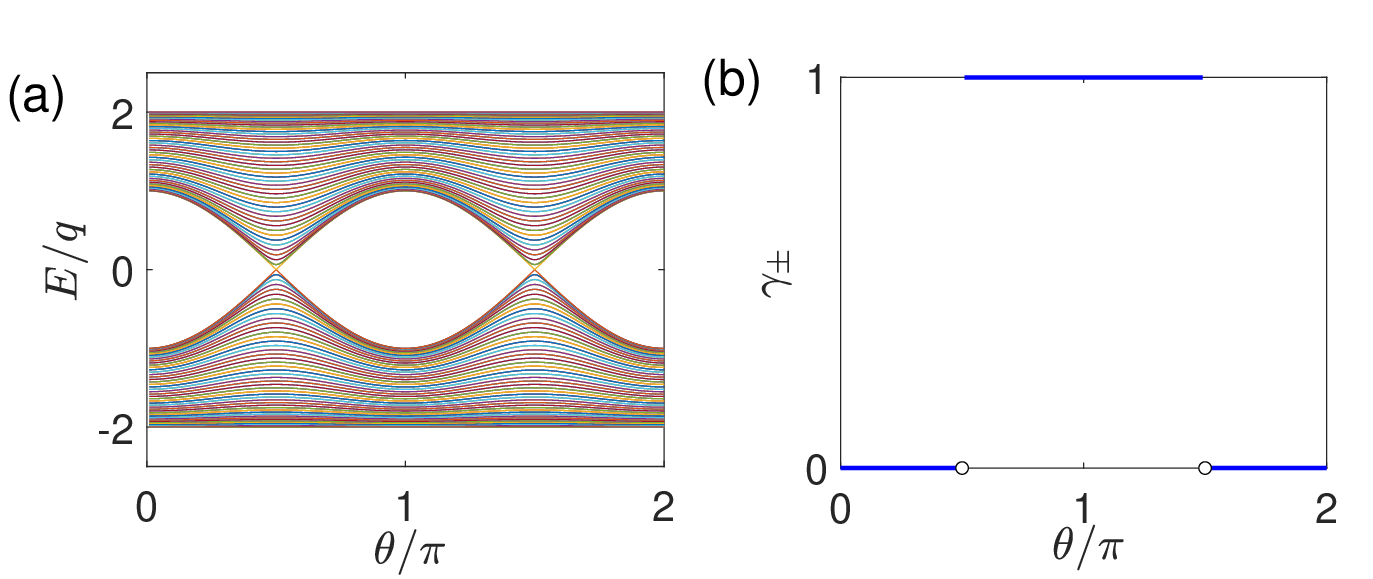}\nonumber\\
   \includegraphics[width=8cm]{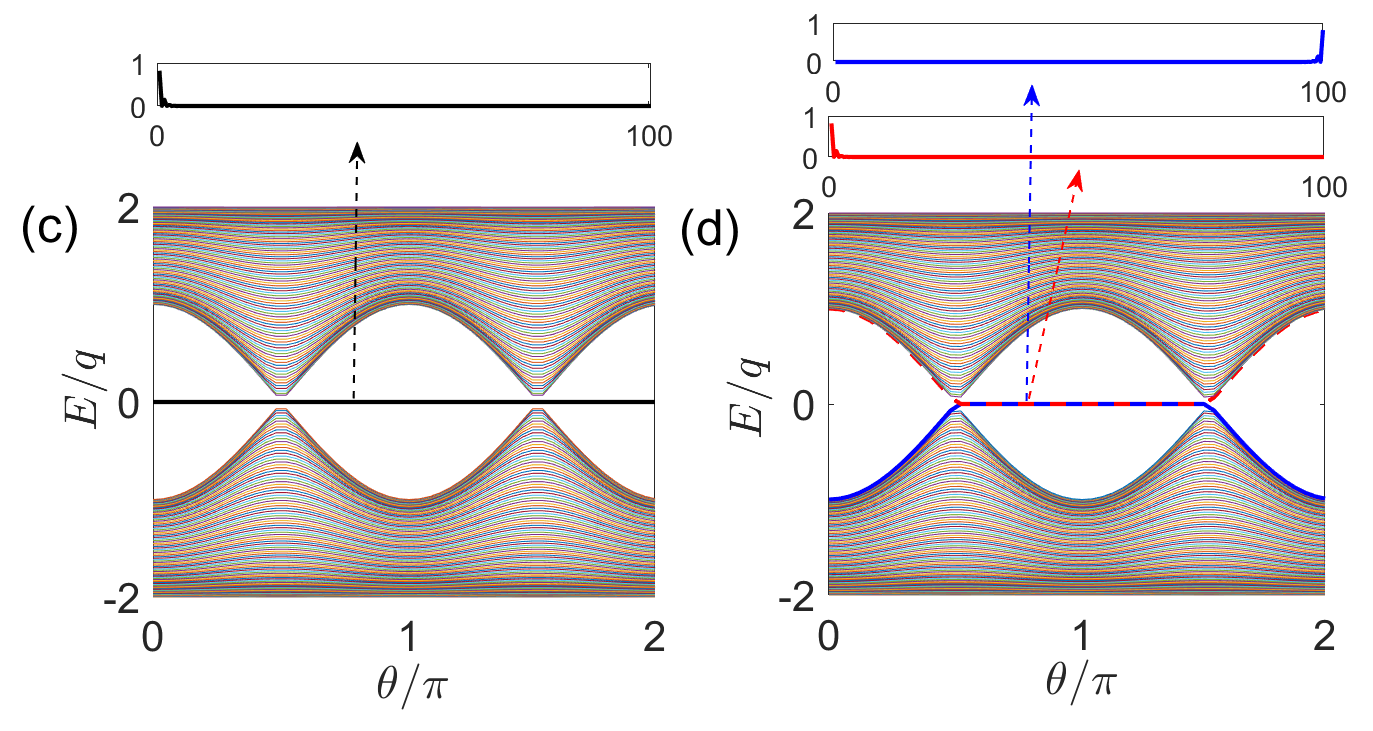}
  \caption{(a) The energy spectrum of the bare SSH model versus $\theta$ for the periodic boundary condition. (b) The winding number versus $\theta$. (c) and (d) The energy spectrum of the SSH model versus $\theta$ for the A-A ($L=101$ for A-B-$\cdots$-B-A setup) and A-B ($L=100$ for A-B-$\cdots$-A-B setup ) open boundary condition, respectively. The parameters are set as $\delta=0.5$.
  The insets in (c) and (d) show the wave function for zero-mode when $\theta=0.8\pi$. }
\label{open}
\end{figure}

The energy spectrum of SSH waveguide with periodical boundary condition is illustrated in Fig.~\ref{open}(a), it shows that the gap is closed and reopened when $\theta$ varies across $\pi/2$ and $3\pi/2$, where the topological phase transition occurs. The topological property of the system can also be characterized by the winding number $\gamma_\pm$, with
 \begin{eqnarray}
\gamma_{\pm}&=&\frac{1}{\pi}\int_{-\pi}^{\pi}dk \left\langle E_{k\pm}\left|i\frac{\partial}{\partial k}\right|E_{k\pm}\right\rangle\nonumber \\
&=&\begin{cases}
1 &  \theta\in(0,\pi/2)\cup(3\pi/2,2\pi)\\
0 &  \theta\in(\pi/2,3\pi/2)
\end{cases},
\end{eqnarray}
which is plotted in Fig.~\ref{open}(b). It implies that the topologically nontrivial phase ($\gamma_\pm=1$) in the regime of $\theta\in(\pi/2,3\pi/2)$ is separated from the topologically trivial phase ($\gamma_\pm=0$).

Alternatively, in the open boundary condition, the topological character of the waveguide is manifested by the boundary states with zero energy.  As shown in Fig.~\ref{open}(c), when the two ends of the chain are the same sublattices, there is only one boundary state
for $0<\theta<2\pi$. On the contrary,  when the two ends of the chain are different sublattices, as shown in Fig.~\ref{open}(d), there are two degenerated boundary states, which locate at the two ends, respectively.

In Sec.~\ref{Theoretical model and energy spectrum} of the main text, we have shown the energy spectrum when the giant atom is coupled to the SSH waveguide under the periodical boundary condition. The energy spectrum is similar to that of the SSH model with open boundary condition as plotted in Figs.~\ref{open} (c) and (d) in this Appendix. {Together with the similarity of the eigen state between our model and that in open SSH chain (as shown in the main text and other parts in the appendix), we find that the giant atom therefore plays a role of an effective boundary.}

\section{Energy Equation}
\label{Energy equation}
In Sec.~\ref{Theoretical model and energy spectrum} of the main text,  we have given the transcendental equations for the eigen energy in Eq.~\eqref{eq:4} and Eq.~\eqref{eq:5} when a giant atom interacts with the waveguide via the A-A or A-B coupling, respectively. Here, we provide the detailed calculations. When the giant atom interacts with the waveguide via the A-A coupling, the Hamiltonian is given by
\begin{equation}
H=H_{\rm SSH}+H_{I,{\rm AA}}.
 \label{HIAA}
\end{equation}
The atom-waveguide coupling Hamiltonian can be also expressed in the momentum space via Fourier transform as
\begin{equation}
H_{I,{\rm AA}}=\frac{g}{\sqrt{L}}\sum_k\left[|e\rangle\langle g|{\hat C}_{A,k}(e^{ikn}+e^{ikm})+{\rm H.c.}\right].
 \label{HIAA(k)}
\end{equation}
In the single-excitation subspace, a general expression of the eigenstate can be given by Eq.~\eqref{psi}.
The Sch\"{o}dinger equation $H|\psi\rangle=E|\psi\rangle$ then leads to
\begin{eqnarray}
EU_e&=&\frac{g}{\sqrt{L}}\sum_{k}A_k(e^{ikn}+e^{ikm}),\nonumber \\
EA_k&=&(t_1+t_2e^{-ik})B_k+\frac{g}{\sqrt{L}}(e^{-ikn}+e^{-ikm})U_e,\nonumber \\
EB_k&=&(t_1+t_2e^{ik})A_k.
 \label{simultaneous equations1}
\end{eqnarray}
Eliminating $A_k, B_k$ and $U_e$ from the above equations, we obtain the transcendental equation for $E$ as
\begin{eqnarray}
E&=&\frac{2g^2}{L}\sum_{k}\left(\frac{E(1+\cos[k(m-n)])}{E^2-\omega_k^2}\right),\nonumber\\
 \label{AAequation}
\end{eqnarray}
then we will obtain Eq.~\eqref{eq:4}.
Obviously, $E=0$ is always the solution. That is, the zero mode for the A-A coupling situation always exists regardless of the value of $\theta$. For the A-B coupling, the Hamiltonian can be written as
\begin{eqnarray}
H_{I,{\rm AB}}
&=&\frac{g}{\sqrt{L}}\sum_k\left[|e\rangle\langle g|({\hat C}_{A,k}e^{ikn}+{\hat C}_{B,k}e^{ikm})+{\rm H.c.}\right],\nonumber \\
 \label{HIAB}
\end{eqnarray}
and the coefficients in Eq.~(\ref{psi}) satisfy the equations
\begin{eqnarray}
EU_e&=&\frac{g}{\sqrt{L}}\sum_{k}(A_ke^{ikn}+B_ke^{ikm}),\nonumber \\
EA_k&=&(t_1+t_2e^{-ik})B_k+\frac{g}{\sqrt{L}}e^{-ikn}U_e,\nonumber \\
EB_k&=&(t_1+t_2e^{ik})A_k+\frac{g}{\sqrt{L}}e^{-ikm}U_e.
 \label{simultaneous equations2}
\end{eqnarray}
As a result, the energy satisfies the equation
\begin{eqnarray}
E=\frac{2g^2}{L}\sum_{k}\frac{E+t_1\cos(kd)+t_2\cos[k(d+1)]}{E^2-\omega_k^2},&\nonumber \\
 \label{ABequation}
\end{eqnarray}
with $d=m-n$, then we will obtain Eq.~\eqref{eq:5}.

\section{Non-trivial Degeneracy  Broken}
\label{Non-trivial degeneracy  broken}
As shown in Fig.~\ref{Energylevel}(d), for the A-B coupling setup, the degeneracy is broken. This is due to the effective coupling, which is induced by the giant atom, between the levels of the original SSH waveguide. To understand this clearly, we adiabatically eliminate the degree of freedom of the atom. To this end, we  first write down the Hamiltonian~\eqref{HIAB} in the eigenspace
\begin{eqnarray}
H_\omega=\sum_k [E_{k+}|E_{k+}\rangle\langle E_{k+}|+E_{k-}|E_{k-}\rangle\langle E_{k-}|],
\end{eqnarray}
\begin{eqnarray}
H_{I,AB}&=&\frac{g}{\sqrt{L}}\sum_k [(\hat{C}_{A,k}e^{ikn}
+\hat{C}_{B,k}e^{ikm})|e\rangle\langle g|+{\rm H.c}]\nonumber \\
&=&\sum[(g_{k+}|G\rangle\langle E_{k+}|+g_{k-}|G\rangle\langle E_{k-}|)|e\rangle\langle g|+{\rm H.c}],\nonumber \\
\end{eqnarray}
where
\begin{eqnarray}
g_{k\pm}=\frac{g}{\sqrt{2L}}(\frac{\omega_ke^{ikn}}{t_1+t_2e^{ik}}\pm e^{ikm}),
\end{eqnarray}
and here we use the relations,
\begin{eqnarray}
\langle E_{k\pm}|\hat{C}^\dagger_{A,k}|G\rangle&=&\frac{\omega_k}{\sqrt{2}(t_1+t_2e^{-ik})}\nonumber \\
\langle E_{k\pm}|\hat{C}^\dagger_{B,k}|G\rangle&=&\pm\frac{1}{\sqrt{2}},
 \label{relations}
\end{eqnarray}
which are obtained by multiplying $\langle E_{k\pm}|$ to Eq.~\eqref{statek0}.

Thus, in the single excitation space, we have
\begin{eqnarray}
\hat{C}^\dagger_{A,k}&=&\frac{\omega_k}{\sqrt{2}(t_1+t_2e^{-ik})}|G\rangle(\langle E_{k+}+\langle E_{k-}|),\nonumber \\
\hat{C}^\dagger_{B,k}&=&\frac{1}{\sqrt{2}}|G\rangle(\langle E_{k+}-\langle E_{k-}|).
\end{eqnarray}

Then  we apply the Schrieffer-Wolff transformation.
In the regime of large detuning $\omega_k\gg |g_{k\pm}|$,  the effective coupling between two modes can be obtained by introducing a unitary transformation $\mathcal {H}=\exp(-\lambda S)H\exp(\lambda S)$, where $S$ is an anti-Hermitian operator. Using the Baker-Hausdorff formula, $\mathcal {H}$ can be written as: (to the second order of ${g_{k\pm}/E_k}$)
\begin{eqnarray}
\mathcal {H}&=&\exp(-\lambda S)(H_{\omega}+H_{I,AB})\exp(\lambda S)\nonumber \\
&=&H_{\omega}+\lambda (H_{I,AB}+[H_{\omega},S])\nonumber \\
&&+\lambda^2([H_{I,AB},S]+\frac{1}{2}[S,[S,H_{\omega}]]),
\end{eqnarray}
where $\lambda$ is a perturbation parameter of the Hamiltonian.
We choose
\begin{eqnarray}
S=\sum_k\left\{\sigma_+(\xi_k|G\rangle\langle E_{k+}|+\chi_k |G\rangle\langle E_{k-}|)-{\rm H.c}\right\},
 \label{S}
\end{eqnarray}
where $\xi_k$ and $\chi_k$ are chosen so that the first order of $\lambda$ to be zero, that is, $H_I + [H_\omega ,S] = 0$, then we have
 \begin{eqnarray}
\xi_k=\frac{g_{k+}}{E_{k+}},
\chi_k=\frac{g_{k-}}{E_{k-}}.
\end{eqnarray}
Since our system is in the large detuning, the giant atom which is initialized as the state $|g\rangle$ will be effectively frozen and
the effective Hamiltonian is obtained as
\begin{eqnarray}
H_{\rm{eff}}&=&\sum_{k,\sigma=\pm}E_{k\sigma}|E_{k\sigma}\rangle\langle E_{k\sigma}|\nonumber \\
&&+\sum_{k,k',(\sigma,\sigma')=\pm}G_{k\sigma,k'\sigma'}|E_{k\sigma}\rangle\langle E_{k'\sigma'}|,
\end{eqnarray}
where the effective coupling strength is $G_{k\sigma,k'\sigma'}=\frac{1}{2}g_{k\sigma}g^*_{k'\sigma'}(\frac{1}{E_{k\sigma}}+\frac{1}{E_{k'\sigma'}})$.

\section{Zero Mode and Bound States}
\label{Zero mode and bound states}
For the giant atom-waveguide coupling system, there exist not only the energy bands but also the states outside the bands. That is, the zero-mode and bound states, which are both the atom-waveguide dressed states. In Sec.~\ref{Sec Zero mode and bound states}, we have shown the results in the topologically nontrivial phase. Here, we will give the detailed derivations in both topologically trivial and nontrivial phases and show the corresponding photon distributions in topologically trivial phase.

\subsection{A-A Coupling}
\label{AA}
For the A-A coupling, Eq.~\eqref{simultaneous equations1} gives
\begin{eqnarray}
\frac{A_k}{U_e}&=&\frac{gE}{\sqrt{L}t_1t_2}(e^{-ikn}+e^{-ikm})f(k),\nonumber \\
\frac{B_k}{U_e}&=&\frac{g(t_1+t_2e^{ik})}{\sqrt{L}t_1t_2}(e^{-ikn}+e^{-ikm})f(k),\nonumber \\
 \label{AAstateL}
\end{eqnarray}
with $f(k)=1/(x-e^{ik}-e^{-ik})$ and $x=(E^2-t_1^2-t_2^2)/t_1t_2$. Performing the Fourier series expansion
\begin{eqnarray}
f(k)=\frac{a_0}{2}+\sum_p\left[\left(\frac{a_p-ib_p}{2}\right)e^{ikp}+\left(\frac{a_p+ib_p}{2}\right)e^{-ikp}\right]\nonumber \\
 \label{fk}
\end{eqnarray}
with
\begin{eqnarray}
a_0&=&\frac{1}{\pi}\int^{\pi}_{-\pi}f(k)dk,\nonumber \\
\frac{a_p+ib_p}{2}&=&\frac{1}{2\pi}\int^{\pi}_{-\pi}f(k)e^{ikp}dk,\nonumber \\
\frac{a_p-ib_p}{2}&=&\frac{1}{2\pi}\int^{\pi}_{-\pi}f(k)e^{-ikp}dk,
\end{eqnarray}
we have
\begin{eqnarray}
f(k)&=&\frac{(-1)^{y+1}}{\sqrt{x^2-4}}\left[1+\sum_{p=1}^L(e^{ikp}a^p+e^{-ikp}a^p)\right],\nonumber \\
 \label{Fourier series expansion}
\end{eqnarray}
where $y=\theta(x)$ is the step function, and $a=(x-\sqrt{x^2-4})/2$ for $x>2$ or  $a=(x+\sqrt{x^2-4})/2$ for $x<-2$. One should note that the above calculations are only valid in the regime of $|x|>2$, that is, the states outside the energy bands. For the states located inside the bands, the discussions are shown below.
\begin{figure}
  \centering
  \includegraphics[width=8cm]{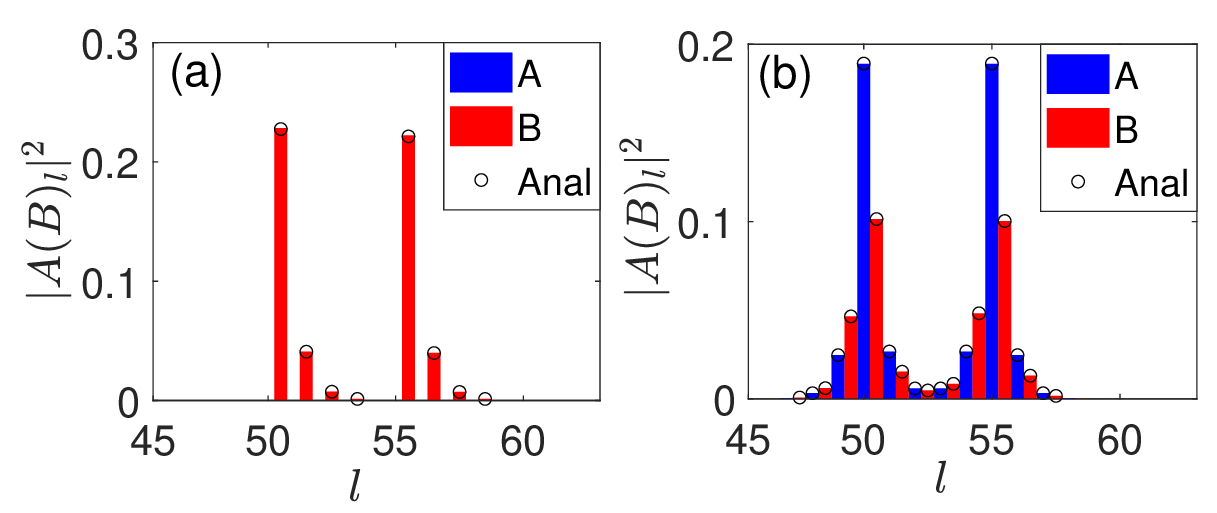}\nonumber\\
  \caption{The photon distribution of the A-A coupling setup for (a) zero-mode and (b)upper bound states. The parameters are set as  {$L=100$}, $n=50, m=55$, $\theta=0.2\pi$, $\delta=0.5q$ and $g=q$.}
\label{boundstate13}
\end{figure}
Substituting the above results into Eq.~(\ref{AAstateL}), we have
\begin{eqnarray}
\frac{A_l}{U_e}&=&\frac{1}{\sqrt{L}}\sum_k e^{ikl} A_k/U_e\nonumber \\
&=&\frac{(-1)^{y+1}T}{\sqrt{x^2-4}}\left(a^{|l-n|}+a^{|l-m|}\right),\label{AAstateA}\\
\frac{B_l}{U_e}&=&\frac{1}{\sqrt{L}}\sum_k e^{ikl} B_k/U_e\nonumber \\
&=&\frac{(-1)^{y+1}}{\sqrt{x^2-4}}[Y_1(a^{|l-n|}+a^{|l-m|})\nonumber \\
&&+ Y_2(a^{|l-n+1|}+a^{|l-m+1|})],\nonumber \\
 \label{AAstateB}
\end{eqnarray}
with $T=gE/(t_1t_2)$ and $Y_r=g/(t_{3-r}),(r=1,2)$. For the zero modes with $E=0$, we will have $x<-2$, thus $a=(x+\sqrt{x^2-4})/2$ is reduced to $(-t_1/t_2)$ for $t_2\geq t_1$ or $(-t_2/t_1)$ for $t_1>t_2$. Then Eq.~\eqref{AAstateA} and  ~\eqref{AAstateB} can be simplified to $A_l=0$ and
\begin{eqnarray}
&&\frac{B_l}{U_e}=\nonumber \\
&&-Y_2\times
\begin{cases}0&(l<n)\\
(-t_2/t_1)^{l-n}&(n\leq l<m)
\\(-t_2/t_1)^{l-n}+(-t_2/t_1)^{l-m}&(m\leq l) \end{cases}\nonumber \\
 \label{AAstateB0}
\end{eqnarray}
for $t_1\geq t_2$ in the topologically trivial phase,
$A_l=0$ and
\begin{eqnarray}
&&\frac{B_l}{U_e}=\nonumber \\
&&Y_2\times\begin{cases} (-t_1/t_2)^{-(l-n)}+(-t_1/t_2)^{-(l-m)}&(l< n)\\
(-t_1/t_2)^{-(l-m)}&(n\leq l<m)\\0&(m\leq l)
\end{cases}\nonumber \\
 \label{AAstateB0}
\end{eqnarray}
for $t_2>t_1$ in the topologically nontrivial phase.

In Sec.~\ref{Sec Zero mode and bound states}, we have shown the photonic distribution for the zero mode in the topologically nontrivial phase. Here, we give the results in the topologically trivial phase in Fig.~\ref{boundstate13}(a), in which we can also observe the chirality.

In addition, the photonic distributions in the bound state for the A-A coupling are shown in Fig.~\ref{boundstate13}(b). Similar to the photon distributions in the topologically non-trivial phase as shown in the main text, it also shows an exponential decay around the atom-waveguide coupling points.

\subsection{A-B Coupling}
Similarly, for the A-B coupling, the amplitudes satisfy
\begin{eqnarray}
\frac{A_k}{U_e}&=&\frac{g}{\sqrt{L}t_1t_2}\left(Ee^{-ikn}+(t_1+t_2e^{-ik})e^{-ikm}\right)f(k),\nonumber \\
&=&\frac{(-1)^{y+1}}{\sqrt{x^2-4}}\left(Ta^{|l-n|}+Y_1a^{|l-m|}+Y_2a^{|l-m-1|}\right),\nonumber \\
\end{eqnarray}
\begin{eqnarray}
\frac{B_k}{U_e}&=&\frac{g}{\sqrt{L}t_1t_2}\left(Ee^{-ikm}+(t_1+t_2e^{ik})e^{-ikn}\right)f(k),\nonumber \\
&=&\frac{(-1)^{y+1}}{\sqrt{x^2-4}}\left(Ta^{|l-m|}+Y_1a^{|l-n|}+Y_2a^{|l-n+1|}\right).\nonumber \\
 \label{ABstateL}
\end{eqnarray}
In this case, the zero mode only exists in the topologically nontrivial phase, that is, $t_2>t_1 (\pi/2<\theta<3\pi/2)$, and the photonic wave function is simplified to
\begin{eqnarray}
\frac{A_l}{U_e}&=&Y_2\begin{cases}(-t_1/t_2)^{(l-m)}&(l> m)\\
0 & (l\leq m)\end{cases} ,
\end{eqnarray}
\begin{eqnarray}
\frac{B_l}{U_e}&=&
Y_2\begin{cases}(-t_1/t_2)^{-(l-n)}&(l< n)\\
0&(l\geq n)\end{cases}.
 \label{SAB1}
\end{eqnarray}

On the other hand, in the topologically trivial phase ($t_1>t_2$), the zero mode disappears, but there exists a single energy level in the gap, and the position of the energy level can be tuned by the size of the giant atom, as discussed in the main text. For this state, the wave function becomes
  \begin{eqnarray}
\frac{A_l}{U_e}&=&
\frac{-1}{\sqrt{x^2-4}}\left(Ta^{|l-n|}+Y_1a^{|l-m|}+Y_2a^{|l-m-1|}\right),\nonumber \\
\end{eqnarray}
  \begin{eqnarray}
\frac{B_l}{U_e}&=&
\frac{-1}{\sqrt{x^2-4}}\left(Ta^{|l-m|}+Y_1a^{|l-n|}+Y_2a^{|l-n+1|}\right),\nonumber \\
 \label{SAB2}
\end{eqnarray}
where $a=(x+\sqrt{x^2-4})/2$, the parameters $T$, $Y_1$ and $Y_2$ are defined in Appendix~\ref{AA}.

\begin{figure}
  \centering
  \includegraphics[width=8cm]{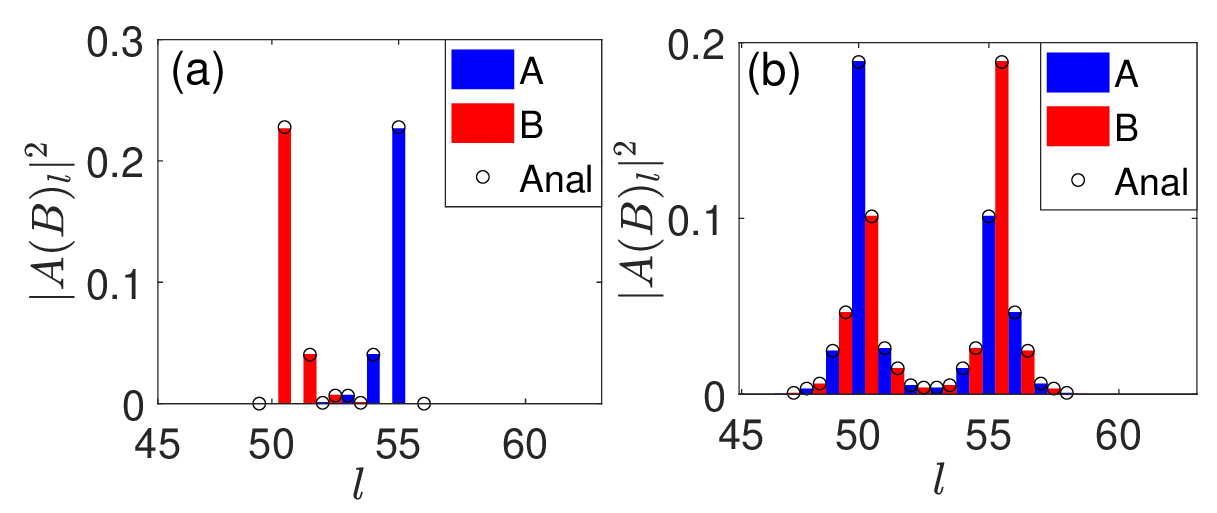}\nonumber\\
  \caption{The photon distribution of the A-B coupling setup for (a)the non-zero energy level in the gap and (b)upper atom-waveguide dressed states. The parameters are set as  {$L=100$}, $n=50, m=55$, $\theta=0.2\pi$, $\delta=0.5q$ and $g=q$.  }
\label{boundstate24}
\end{figure}

As shown in Fig.~\ref{boundstate24}(a), where we plot the photonic distributions in the topologically trivial phase,  the photon is occupied in the A(B) sublattice which is on the right(left) half side inside the regime covered by the giant atom for the energy level in the gap. Alternatively, in the topologically non-trivial phase, as shown in Fig.~\ref{Zmode}(b), there is photonic excitation in the B (A) sublattice on the left (right) side of the giant atom, but there is no occupation inside the atom. Therefore, the chirality of the photonic distributions witness the topological phase transition when a giant atom interacts with the SSH waveguide via the A-B coupling. Besides, the upper bound states in the topologically trivial phase are also shown in Fig.~\ref{boundstate24}(b), which has a similar symmetry as that in the topologically nontrivial phase (see, the main text).

In  addition, we would like to compare the zero modes in the bare SSH waveguide with open boundary condition and the giant atom-waveguide coupling system with periodical boundary condition. In the open boundary condition, the wave functions of zero modes are given in Figs.~\ref{open}(c) and (d), they show that the photon occupies the sites nearby the edge, which is therefore usually named as edge state. For the atom-waveguide coupling system, as shown in the main text, the photon is excited near the giant atom. In this sense, the giant atom plays a role of effective boundary.

\section{Robustness of Zero Modes}
\label{Robustness of zero modes}
In this section, we numerically study the robustness of the zero modes to the disorder, atomic dissipation, asymmetric atom-waveguide coupling and the next-nearest-neighbor hopping in the SSH waveguide.

\subsection{Disorder}
\begin{figure}
  \centering
  \includegraphics[width=8cm]{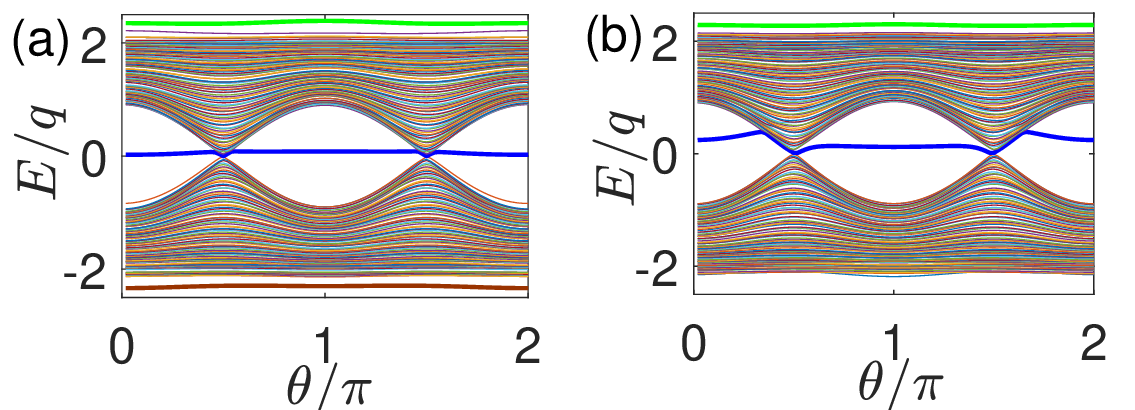}\nonumber\\
  \includegraphics[width=8cm]{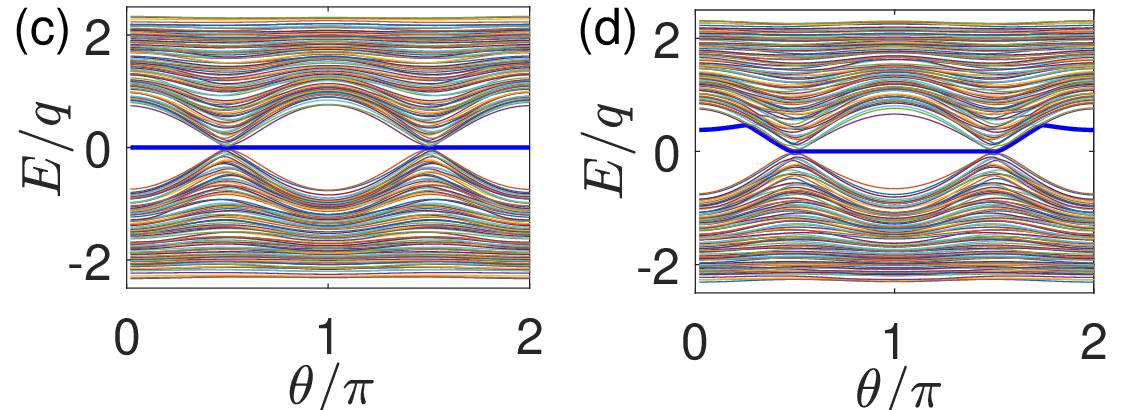}\nonumber\\
  \caption{The energy spectrum versus $\theta$ in via the A-A coupling in (a) and (c) or the A-B coupling in (b) and (d), under the periodical boundary condition. The disorders are set up by onsite $\omega$ in (a) and (b), the hopping $t_1$, $t_2$ in (c) and (d).
  The parameters are set as $\Omega=\omega$=0, $\delta=0.5, L=100, n=50, m=51, g=q, \mu=0$ and $\sigma=0.1938$.}
\label{disorder1}
\end{figure}

\begin{figure}
  \centering
  \includegraphics[width=8cm]{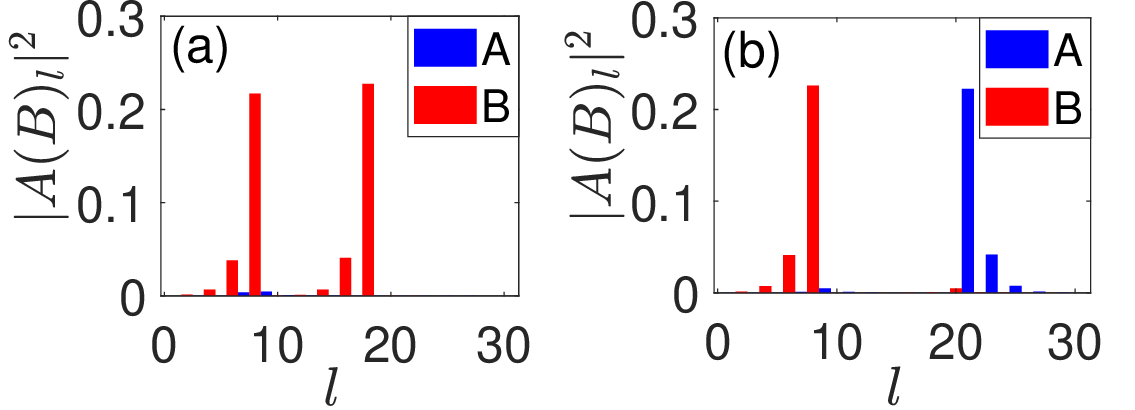}\nonumber\\
    \includegraphics[width=8cm]{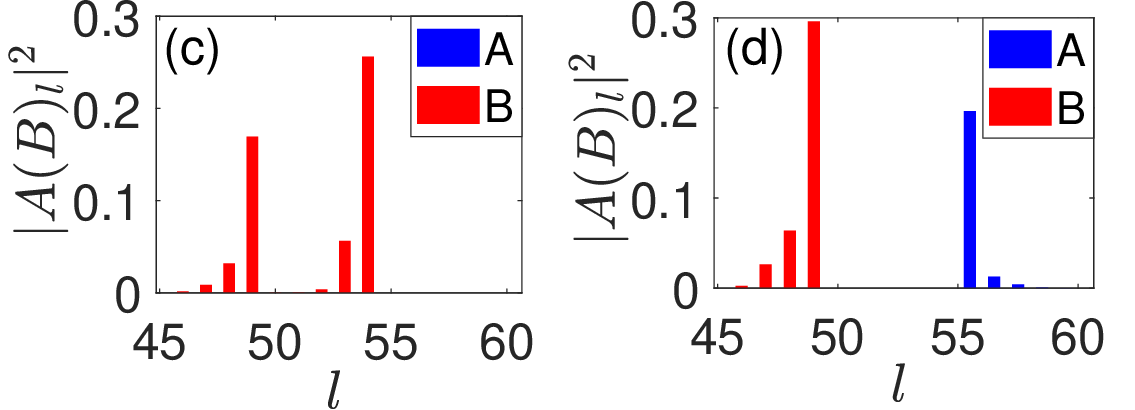}\nonumber\\
  \caption{ The photon distribution in zero mode with the A-A coupling in (a) and (c) or the A-B coupling in (b) and (d).
The disorders are set up by onsite $\omega$ for (a) and (b), hopping $t_1$, $t_2$ for (c) and (d).
  The parameters are set as $\Omega-\omega=0$, $\delta=0.5, L=100, n=50, m=55, g=q, \mu=0$ and $\sigma=0.1938$.}
\label{disorder2}
\end{figure}

We first consider the effect of disorder. The disorder comes from two aspects, one is the on site frequency disorder, while the other one is the hopping disorder~\cite{TV}.   We consider a random disorder with standard Gaussian distribution, with central value $\mu$ and full width at half maximum $\sigma$.  The energy spectrum is given in Fig.~\ref{disorder1}, where (a) and (b) represent the on site frequency disorder, while (c) and (d) represent the hopping disorder. The corresponding wave function is given in Fig.~\ref{disorder2}.

As shown in Fig.~\ref{disorder1}, the zero modes always exist in both kinds of disorders. For the onsite frequency disorder, the wave function of zero-mode is only slightly perturbed as shown in Figs.~\ref{disorder2}(a) and (b), but the symmetry character is broken for the hopping disorder as shown in Figs.~\ref{disorder2}(c) and (d). {Here, we use the phrase ''symmetry'' to denote the character of the photonic distributions in the waveguide, but not the physical nature of the system, i.e., the Hamiltonian.} Meanwhile, both of the spectrum and the wave function of the bound states outside the bands are not robust to both of two kinds of disorder.

\subsection{Dissipation}

We now consider the spontaneous emission of the giant atom by phenomenologically introducing a non-Hermitian Hamiltonian,
\begin{eqnarray}
H=H_{\rm SSH}+H_{I,{\rm AA(AB)}}-i\gamma_e|e\rangle\langle e|.
 \label{dissipation}
\end{eqnarray}
where $\gamma_e$ is the decay rate of the atom.
\begin{figure}
  \centering
  \includegraphics[width=8cm]{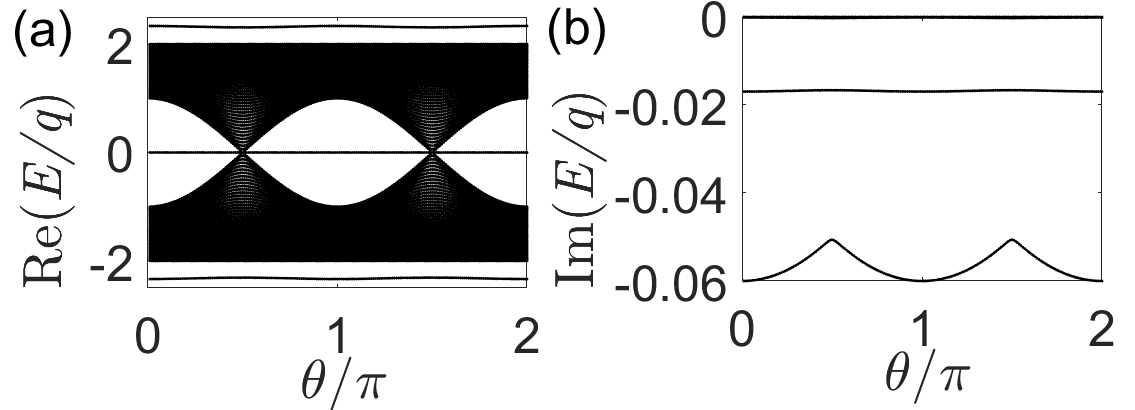}\nonumber\\
  \includegraphics[width=8cm]{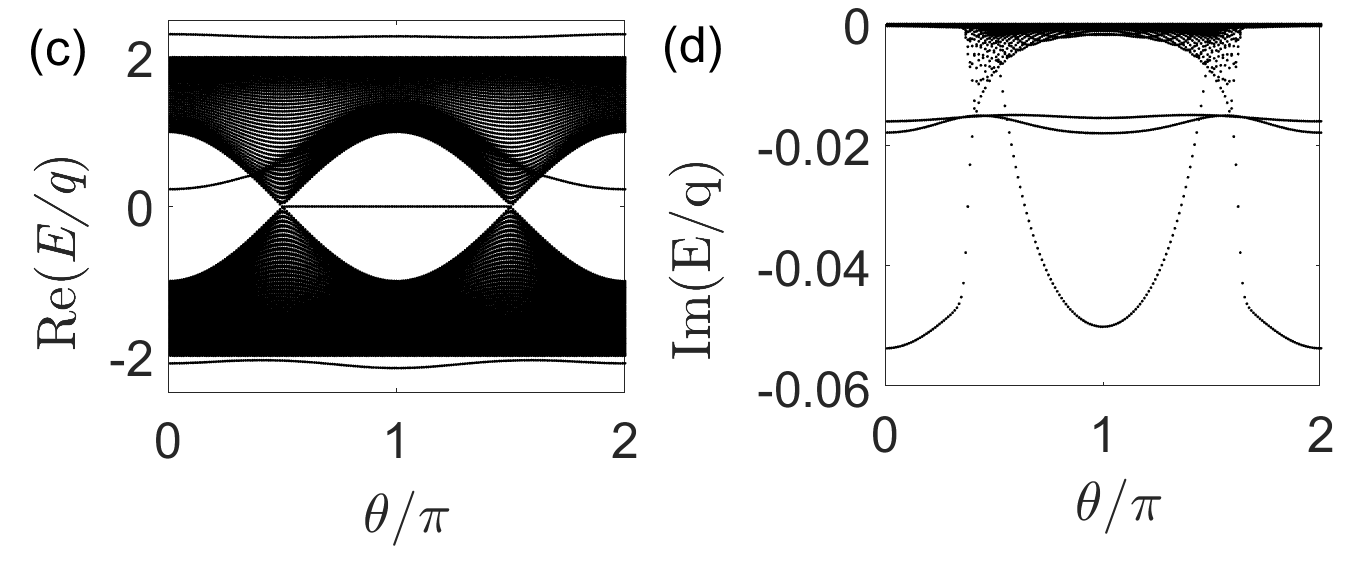}\nonumber\\
  \caption{The real (imaginary) energy spectrum of SSH chain coupling a giant atom via the A-A in (a) and (b) or the A-B coupling in (c) and (d).
  The parameters are set as  $\delta=0.5, L=100, g=q, \gamma_e=0.1q, n=50$ and $m=51$.}
\label{diss}
\end{figure}
As shown in Fig.~\ref{diss},  we plot the real and imaginary parts of the energy spectrum of the system with the A-A and A-B couplings, respectively. Compared with Figs.~1(b) and (c) in the main text and Figs.~\ref{diss}(a) and (c), we find that the real part of the energy spectrum is nearly unchanged, we can still observe the topological nature, that is, the zero modes. Therefore, the topology of the giant atom-waveguide system is robust to the atomic dissipation. Meanwhile, we can also observe some exotic effects in the imaginary part of the energy spectrum as shown below.

For the A-A coupling, as shown in Fig.~\ref{diss}(b), the imaginary part of the energy spectrum is zero except two isolated energy levels. One exists around ${\rm Im}(E/q)=0.02q$ while the other one oscillates in the region of ${\rm Im}(E/q)=-(0.05-0.06)q$. Furthermore, the oscillating curve reaches the maximum value at $\theta=\pi/2$ and $\theta=3\pi/2$, which implies a topological phase transition.

For the A-B coupling, the imaginary part of the energy spectrum has different structures in  the  topologically non-trivial and topologically trivial phases as shown in Fig.~\ref{diss} (d).
There are three separated energy levels,  two of them locate around ${\rm Im}(E/q)=-0.02q$ in both of the topologically trivial and nontrivial phases, and intersect with each other at $\theta=\pi/2$ and $\theta/2=3\pi/2$, where the phase transition occurs.  The third one, however, nearly locates around ${\rm Im}(E/q)=-0.05q$ in the topologically trivial phase but behaves quadratically in the topologically nontrivial phase. Except for these three isolated levels, the imaginary part of the energy spectrum is always zero in the topologically trivial phase and acquires continual but small values in the topologically nontrivial phase.

\subsection{Asymmetric Coupling and Next-nearest-neighbor Hopping}

\begin{figure}
  \centering
  \includegraphics[width=8cm]{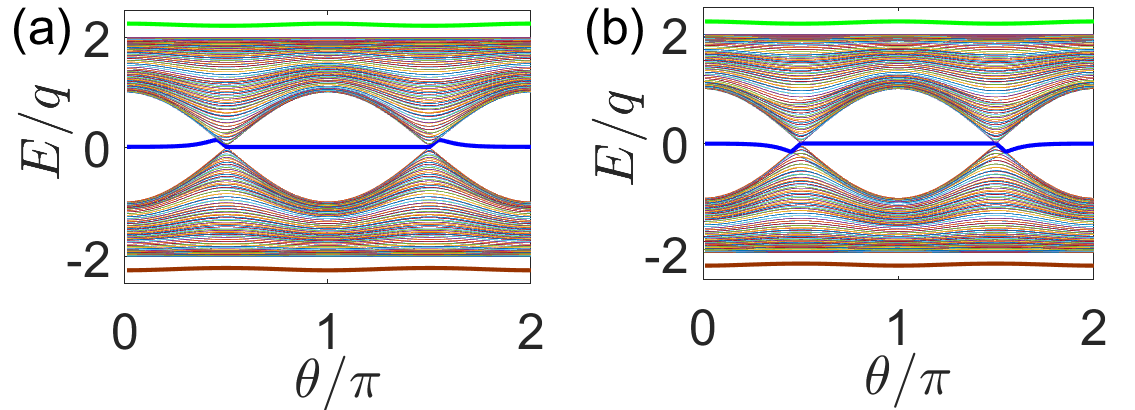}\nonumber\\
  \includegraphics[width=8cm]{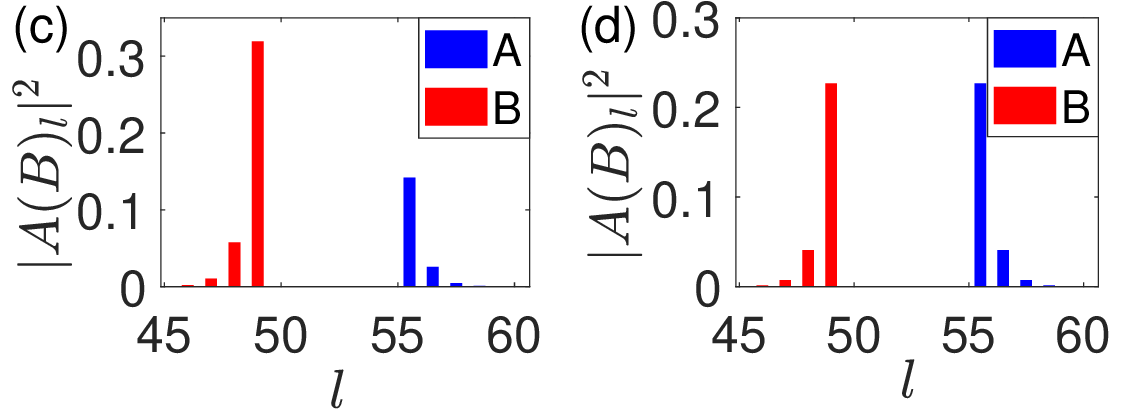}\nonumber\\
  \caption{(a) and (b) The energy spectrums for the SSH model coupling with the giant atom via the asymmetric A-B coupling.  (c) and (d) The zero-mode eigenstates for the SSH model coupling with the giant atom via the asymmetric A-B coupling. The parameters are set as $\delta$=0.5, $L=100$. $\theta=0.8\pi$ for (c) and (d). $g_1=1.2q$ and $g_2=0.8q$ for (a) and (c). $g_1=q$ and $g_2=-q$ for (b) and (d).}
\label{asymmetric}
\end{figure}

We now consider the asymmetric coupling between the two atom-waveguide coupling points. In this case, the interaction Hamiltonian of the system can be written as
 \begin{eqnarray}
H_{I,{\rm AA(AB)}}=g_1\sigma_+\hat{C}_{A,n}+g_2\sigma_+\hat{C}_{A(B),m}+{\rm H.c.}.
 \label{dissipation}
\end{eqnarray}
where $g_1\neq g_2$.
In Figs.~\ref{asymmetric}(a) and (b) [(c) and (d)], we plot the energy spectra (zero-mode eigenstates) with different coupling strengths $g_1$ and $g_2$ for the A-B coupling. With the parameters $g_1=1.2q$ and $g_2=0.8q$, the energy spectrum in Fig.~\ref{asymmetric}(a) keeps the original nature. However, compared with the symmetry coupling, the zero-mode eigenstate in Fig.~\ref{asymmetric}(c) shows an asymmetric character, which means that the photon occupation amplitude in each site depends on $g_1$ and $g_2$.

For other parameters $g_1=q$ and $g_2=-q$, we obtain the energy spectrum in Fig.~\ref{asymmetric}(b), which exhibits a negative energy in the gap. It thus supplies another approach to adjust the nonzero energy in the gap besides tuning $d=|m-n|$ (as shown in Fig.~\ref{ABlevel}).  For the wave function, as shown in Fig.~\ref{asymmetric}(d), it maintains the chiral and symmetry character, that is, the photonic population in B(A) sublattice on the left(right) side of the giant atom possess the same amplitudes.

For the A-A coupling, the zero mode always exist for $\theta\in(0,2\pi)$, with the asymmetric atom-waveguide coupling. Also, similarly to the A-B coupling, the symmetry photonic distributions are broken as long as $|g_1|\neq|g_2|$.

\begin{figure*}
  \centering
  \includegraphics[width=15cm]{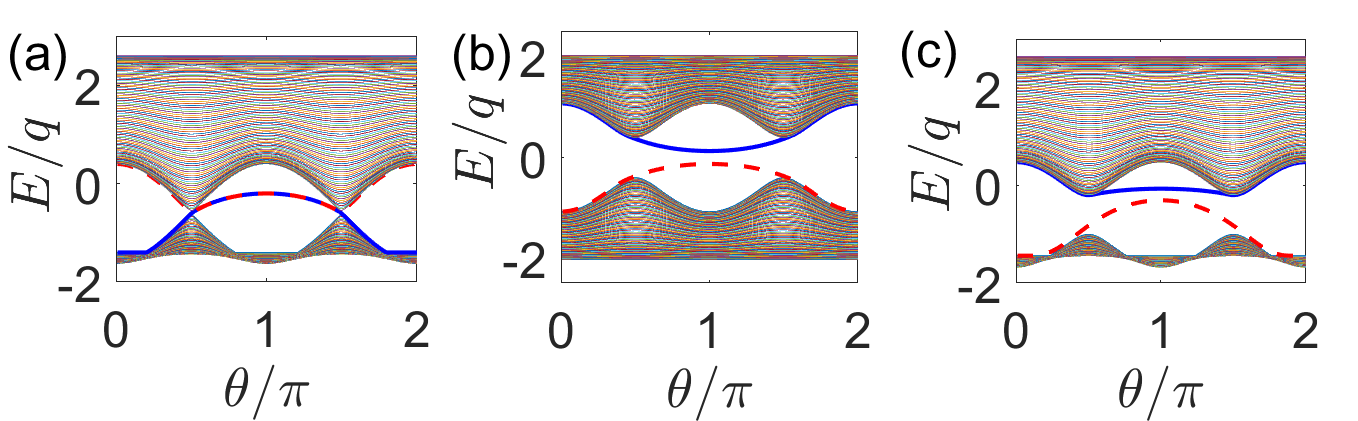}\nonumber\\
    \includegraphics[width=15cm]{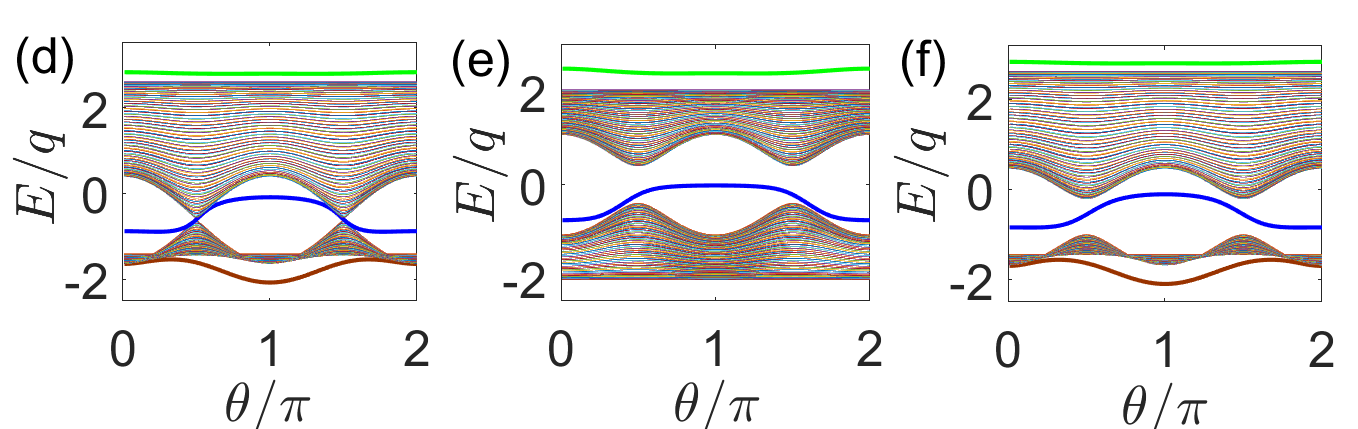}\nonumber\\
        \includegraphics[width=15cm]{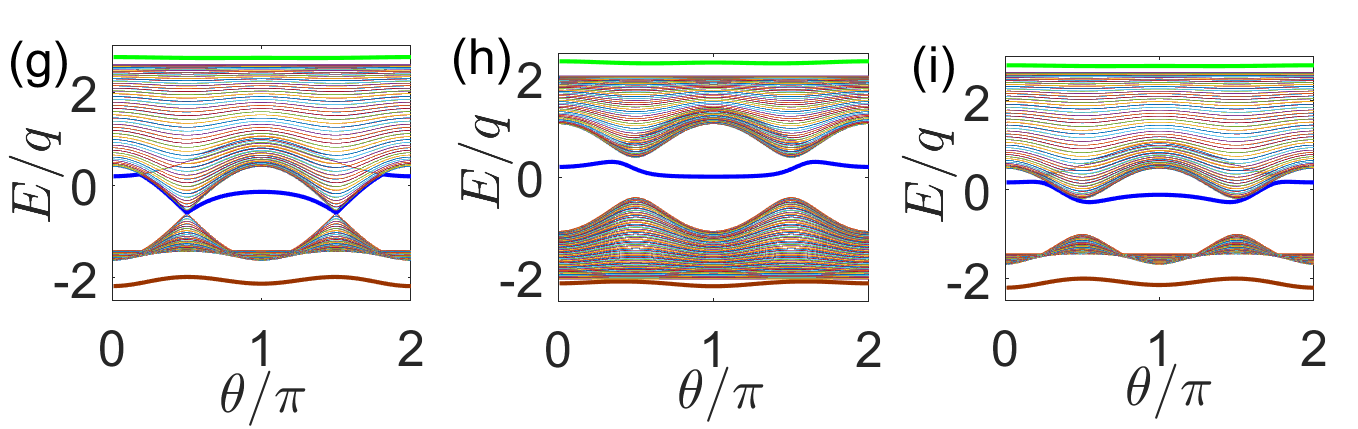}\nonumber\\
  \caption{(a), (b) and (c) The energy spectrum for SSH model for the open boundary with NNN coupling. (d), (e), (f), (g), (h) and (i) The energy spectrum for extended cyclical SSH model coupling with the giant atom via A-B coupling. The parameters are set as $\delta=0.5q, L=100, g=q$. (d), (e) and (f): $n=50, m=50$.  (g), (h) and (i): $n=50, m=51$. The next nearest coupling strength: $t_A=0.3q, t_B=0.3q$ for (a), (d) and (g). $t_A=0.2q, t_B=-0.2q$ for (b), (e) and (h). $t_A=0.5q, t_B=0.1q$ for (c), (f) and (i).}
\label{nextnearest}
\end{figure*}

In addition, we consider the effect of next-nearest-neighbor (NNN) hopping. The Hamiltonian with the NNN hopping is $H'=H+H_{\rm NNN}$ where
 \begin{eqnarray}
H_{{\rm NNN}}=\sum_l[t_{A}\hat{C}^{\dag}_{A,l}\hat{C}_{A,l+1}+t_{B}\hat{C}_{B,l}\hat{C}_{B,l+1}+{\rm H.c.}].\nonumber \\
 \label{dissipation}
\end{eqnarray}
To show how the NNN hopping terms affect the properties of the SSH model, we first plot the energy spectrum in Figs.~\ref{nextnearest} (a), (b) and (c) on the open boundary condition without the giant atom~\cite{LLi}. The zero-modes of the SSH model are protected by both the inversion symmetry and particle-hole symmetry. These symmetries can be broken by the NNN hopping terms. For example, when $t_A=t_B$, the particle-hole symmetry is broken but the inversion symmetry is retained. As a result, the zero mode is shifted lower but the gap is not opened.  With the same NNN hopping, we plot the energy spectra for the A-B coupling on the cycle condition in Figs.~\ref{nextnearest}(d) for $m-n=0$  and (g) for $m-n=1$.

For another set of parameters ($t_A=-t_B=0.2q$), we obtain the energy spectra in Fig.~\ref{nextnearest}(b), which shows a symmetric character for the upper and lower energy bands and the gap is opened. The zero-modes are split and are laterally zygomorphic.
Introducing the giant atom for the A-B coupling, we can still obtain the same symmetry in the two bands as shown in Figs.~\ref{nextnearest}(e) and (h). However, the zero-mode and bound state show an asymmetrical structure.

At last, we set $t_A=0.5q$ and $t_B=0.1q$, in which both the particle-hole symmetry and the inversion symmetry are broken and the energy spectrum is plotted in Fig.~\ref{nextnearest}(c). As a comparison, we also give the spectrum with the A-B coupling in Figs.~\ref{nextnearest}(f) and (i), which are respectively similar to those in (d) and (g), except the open of the energy gap.

\section{Emission of Giant Atom}
\label{Emission of giant atom}
\subsection{Temperature Effect}
\label{Temperature effect}
In the viewpoint of quantum open system, the SSH waveguide serves as a topological environment of the giant atom, and the atom will undergo the emission process. To demonstrate the emission dynamics of the giant atom, we resort to the master equation approach.
Generally speaking, the master equation for a quantum open system can be written as~\cite{openbook}
 \begin{eqnarray}
\frac{d\rho_S(t)}{dt}=-{\rm Tr}_{R}\int_{0}^{\infty}[V^{I}(t),[V^{I}(t'),\rho_S(t)\otimes\rho_{R}]]dt',\nonumber \\
\label{motion4}
\end{eqnarray}
where $V^{I}(t)$ is the system-environment interaction Hamiltonian in the interaction picture, $\rho_S$ is the reduced density matrix of the considered open system and $\rho_R$ is the density matrix of the environment, which is considered as time independent, i.e., $\rho_R\equiv\rho_R(0)$. In the interaction picture, the interaction Hamiltonian between the giant atom and the SSH waveguide can be written as
\begin{eqnarray}
V^{I}_{\rm AA}(t)
&=&\frac{g}{\sqrt{L}}\sum_k[(e^{ikn}+e^{ikm})\frac{\omega_k}{\sqrt{2}(t_1+t_2e^{ik})}\times\nonumber \\
&&e^{-i(\omega+\omega_k-\Omega)t}\sigma^{+}|G\rangle\langle E_{k+}|\nonumber \\
&&+(e^{ikn}+e^{ikm})\frac{\omega_k}{\sqrt{2}(t_1+t_2e^{ik})}\times\nonumber \\
&&e^{-i(\omega-\omega_k-\Omega)t}\sigma^{+}|G\rangle\langle E_{k-}|+{\rm H.c.}],
\end{eqnarray}
\begin{eqnarray}
V^{I}_{\rm AB}(t)
&=&\frac{g}{\sqrt{L}}\sum_k[(\frac{e^{ikn}\omega_k}{\sqrt{2}(t_1+t_2e^{ik})}+\frac{e^{ikm}}{\sqrt{2}})\times\nonumber \\
&&e^{-i(\omega+\omega_k-\Omega)t}\sigma^{+}|G\rangle\langle E_{k+}|\nonumber \\
&&+(\frac{e^{ikn}\omega_k}{\sqrt{2}(t_1+t_2e^{ik})}-\frac{e^{ikm}}{\sqrt{2}})\times\nonumber \\
&&e^{-i(\omega-\omega_k-\Omega)t}\sigma^{+}|G\rangle\langle E_{k-}|+{\rm H.c.}],
\label{ABin}
\end{eqnarray}
for the A-A coupling and the A-B coupling, respectively. We consider that the waveguide, which acts as an effective bath, is furthermore surrounded by an thermal bosonic bath with temperature $T$ satisfying $\hbar\Omega/(k_BT)\gg1$, then the photon number inside the waveguide is no more than $1$, that is, we only consider the zero and single photon state. As a result, due to the bosonic distribution, the probability for zero photon excitation is
\begin{eqnarray}
P_G&=&{\rm Tr}_R[\rho_R|G\rangle\langle G|]\nonumber \\
&=&\prod_{k'}\left(1-e^{-\hbar (\omega+\omega_{k'})/(k_{B}T)}\right)\times\nonumber \\
&&\left(1-e^{-\hbar (\omega-\omega_{k'})/(k_{B}T)}\right),
\label{PG}
\end{eqnarray}
and the probabilities to find a thermal photon in the state $E_{k_{\pm}}$ are
 \begin{eqnarray}
P_{E_{k_{\pm}}}=P_Ge^{-\hbar (\omega\pm\omega_k)/(k_{B}T)}.
\label{PEk+}
\end{eqnarray}
Substituting the above results into Eq.~(\ref{motion4}), we have
\begin{widetext}
 \begin{eqnarray}
\frac{d}{dt}\rho_S(t)&=&-P_G\sum_k\{[(\alpha_k+\beta_k)\sigma^{+}\sigma^{-}\rho_{S}(t)+(\alpha_k^{\ast}+\beta_k^{\ast})\rho_{S}(t)\sigma^{+}\sigma^{-}
-(\alpha_k+\beta_k+\alpha_k^{\ast}+\beta_k^{\ast})\sigma^{-}\rho_{S}(t)\sigma^{+}]\nonumber \\
&&+n_{k+}[\alpha^{\ast}_k\sigma^{-}\sigma^{+}\rho_{S}(t)+\alpha_k\rho_{S}(t)\sigma^{-}\sigma^{+}
-(\alpha_k+\alpha^{\ast}_k)\sigma^{+}\rho_{S}(t)\sigma^{-}]\nonumber \\
&&+n_{k-}[\beta^{\ast}_k\sigma^{-}\sigma^{+}\rho_{S}(t)+\beta_k\rho_{S}(t)\sigma^{-}\sigma^{+}
-(\beta_k+\beta^{\ast}_k)\sigma^{+}\rho_{S}(t)\sigma^{-}]\},
\label{motion5}
\end{eqnarray}
\end{widetext}
with $n_{k\pm}=e^{-\hbar (\omega\pm\omega_k)/(k_{B}T)}$ and
\begin{eqnarray}
\alpha_k&=&\frac{g^2\pi}{L}\left\{1+\cos(kd)\right\}\delta(\omega+\omega_k-\Omega),\\
\label{alpha}
\beta_k&=&\frac{g^2\pi}{L}\left\{1+\cos(kd)\right\}\delta(\omega-\omega_k-\Omega),
\label{beta}
\end{eqnarray}
for the A-A coupling and
 \begin{eqnarray}
\alpha_k=
 &&\frac{g^2\pi}{L}\left\{1+\frac{t_1\cos(kd)+t_2\cos[k(d-1)]}{\omega_k}\right\}\nonumber \\
 &&\times\delta(\omega+\omega_k-\Omega),
  \end{eqnarray}
  \begin{eqnarray}
 \beta_k=
 &&\frac{g^2\pi}{L}\left\{1-\frac{t_1\cos(kd)+t_2\cos[k(d-1)]}{\omega_k}\right\}\nonumber \\
&& \times\delta(\omega-\omega_k-\Omega),
\label{beta}
 \end{eqnarray}
for the A-B coupling. In the above equations, we have defined $d=n-m$.

\begin{figure*}
  \centering
  \includegraphics[width=15cm]{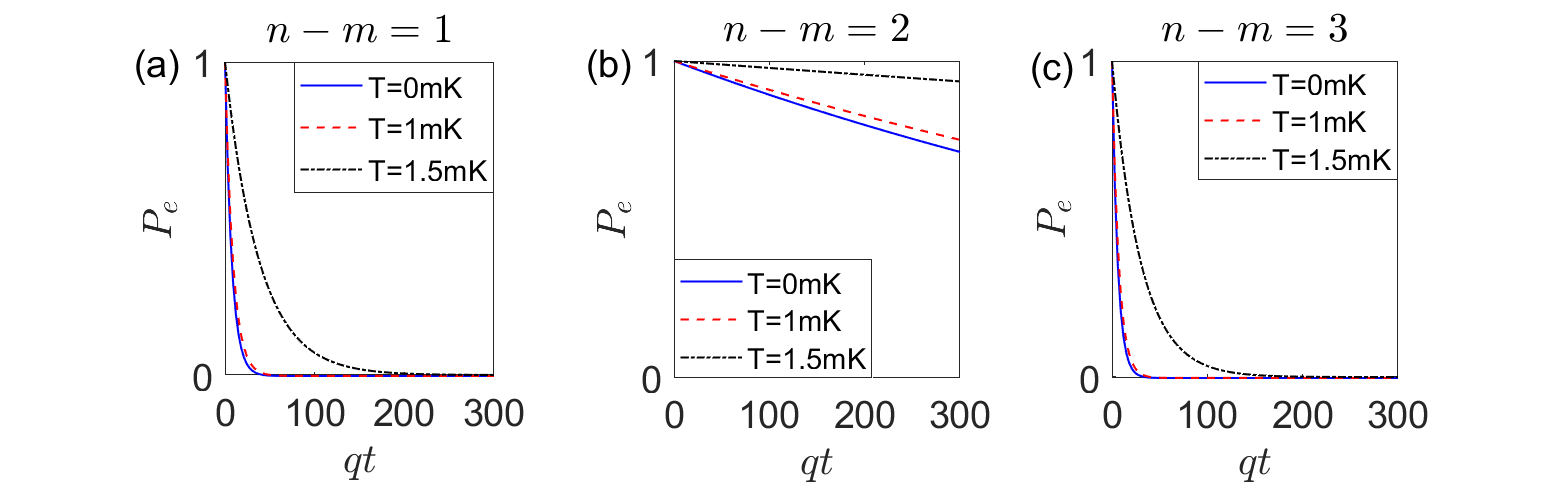}\nonumber\\
  \caption{Evolution of the giant atom for the A-A coupling  with different temperatures.
  The parameters are set as ($\Omega-\omega)/q=1.5, \delta=0.5, L=100, g=0.2q$, $\omega=10^{9}$Hz, $q=0.1\omega$ and $\theta=0.8\pi$.}
\label{temAA}
\end{figure*}

\begin{figure*}
  \centering
  \includegraphics[width=15cm]{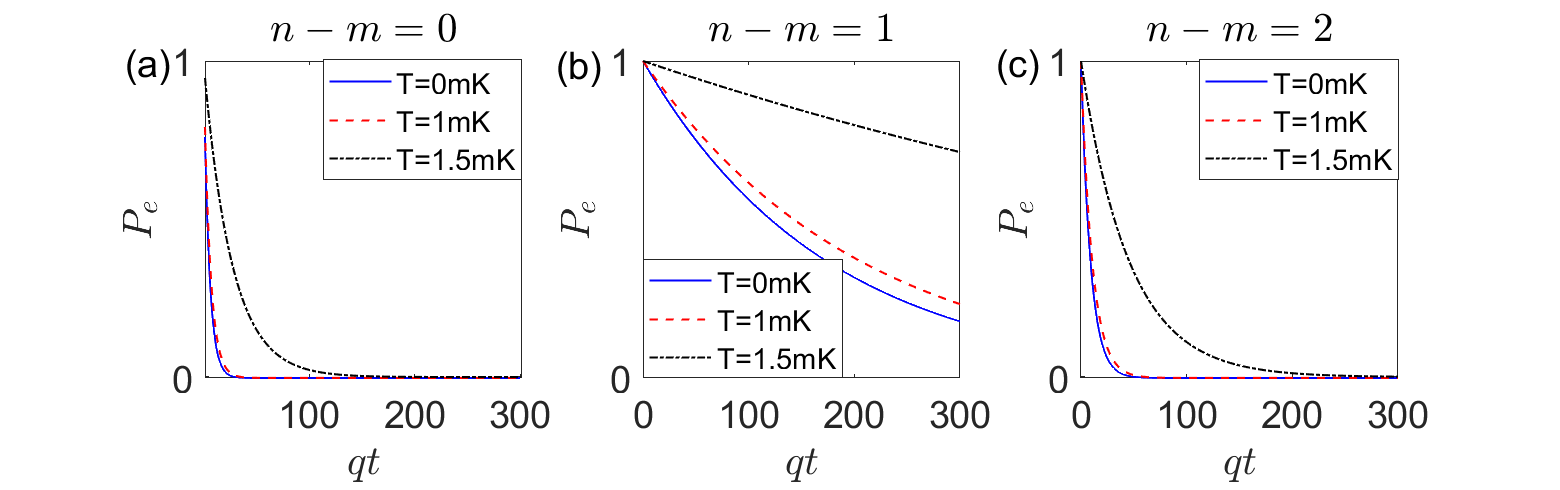}\nonumber\\
  \includegraphics[width=15cm]{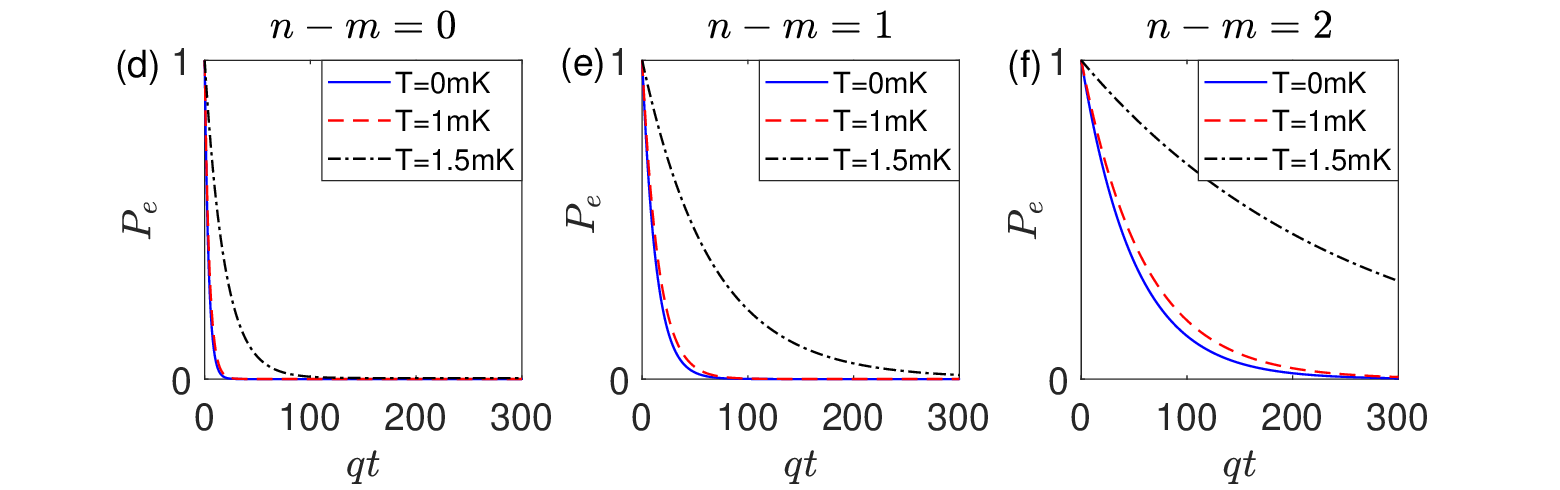}\nonumber\\
  \caption{Evolution of the giant atom for the A-B coupling with different temperatures.
  The parameters are  ($\Omega$-$\omega$)/q=1.5, $\delta=0.5, L=100, g=0.2q, \omega=10^9Hz$, $q=0.1\omega$ and $\theta=0.8\pi$ for (a), (b), (c) and $\theta=0.2\pi$ for (d), (e), (f).}
\label{temAB}
\end{figure*}

In Sec.~\ref{Non-Markovian retardation effect} of the main text, we have shown the atomic dynamics at zero temperature.  Now, let us study the effect of temperature. In Figs.~\ref{temAA} and ~\ref{temAB}, we plot the population of the giant atom in the excited state in different temperatures for $|n-m|=0, 1$ and $2$ based on the master equation, that is, we neglect the non-Markovian effect here. For the A-A coupling, we find that the dynamics is similar in the topologically trivial and non-trivial phases. Therefore, we only give the results in the topologically non-trivial phase in Fig.~\ref{temAA}. The results for the A-B coupling are given in Fig.~\ref{temAB} in topologically trivial and non-trivial phases, respectively. As shown in Figs.~\ref{temAA} and ~\ref{temAB}, in the low temperature (that is, $k_B T/(\hbar\omega)<1$), the dynamics is nearly independent of the temperature. In fact, in such a low temperature, the SSH waveguide is nearly in the vacuum state. However, as the increase of the temperature, the waveguide acquires the thermal photonic excitation, and the emission of the giant atom is suppressed a lot.

\subsection{Non-Markovian Retardation Effect}
\label{Non-Markovian Retardation effect}
{In the above subsection, we have investigated the emission of the giant atom under the Markovian approximation and the non-Markovian effect at zero temperature is furthermore demonstrated in Sec.~\ref{Non-Markovian retardation effect} (see Fig.~\ref{markovian2}) of the main text. From the view of atomic population dynamics, the non-Markovian effect is exhibited by the Rabi oscillation and retardation effect when the frequency of the giant atom is located nearby the edge of the band and well-inside the band of the waveguide, respectively.}

\begin{figure}
  \centering
  \includegraphics[width=8cm]{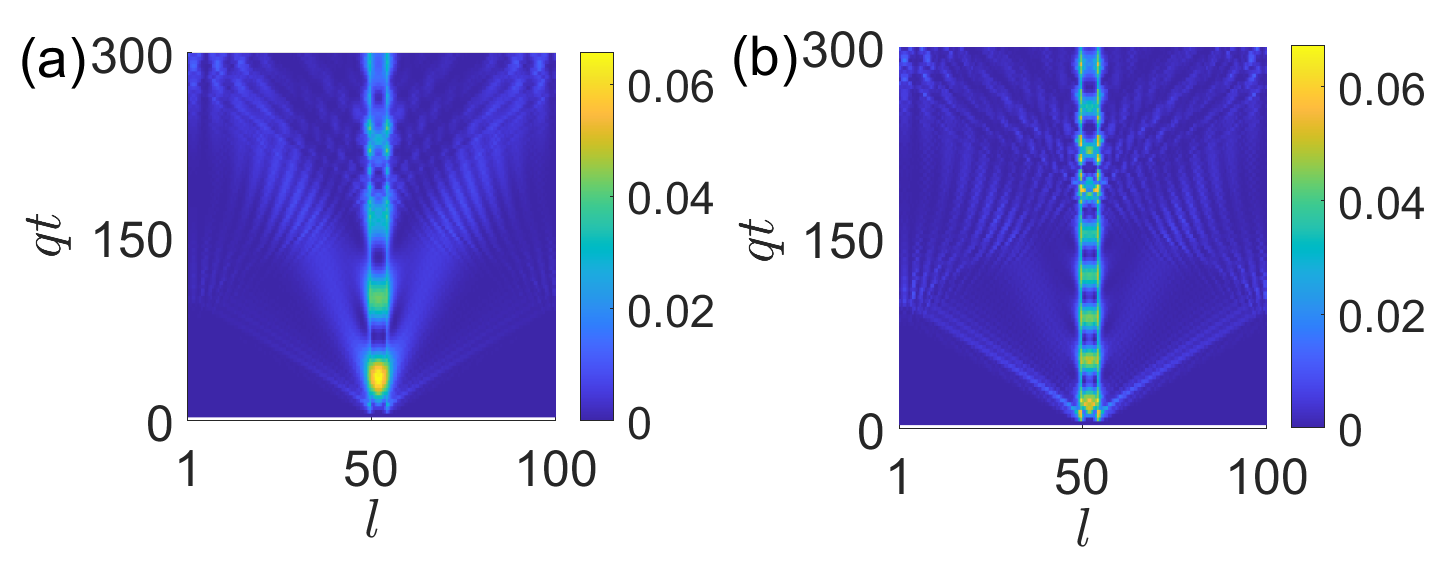}
  \includegraphics[width=8cm]{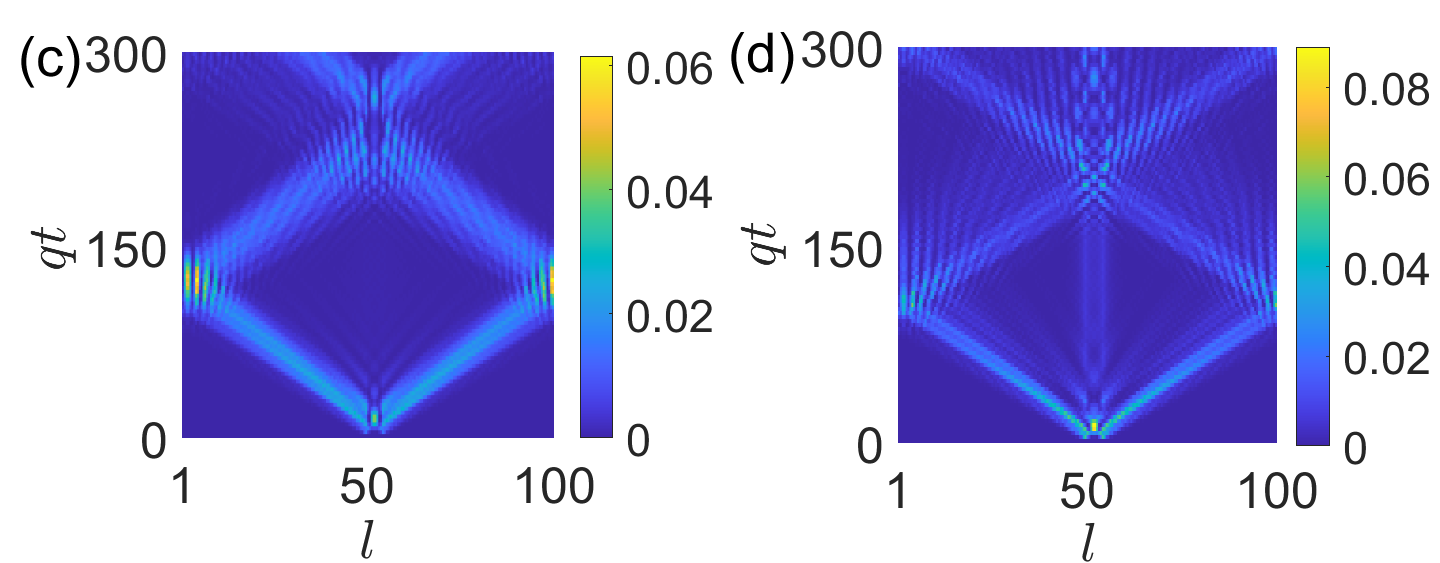}
  \includegraphics[width=8cm]{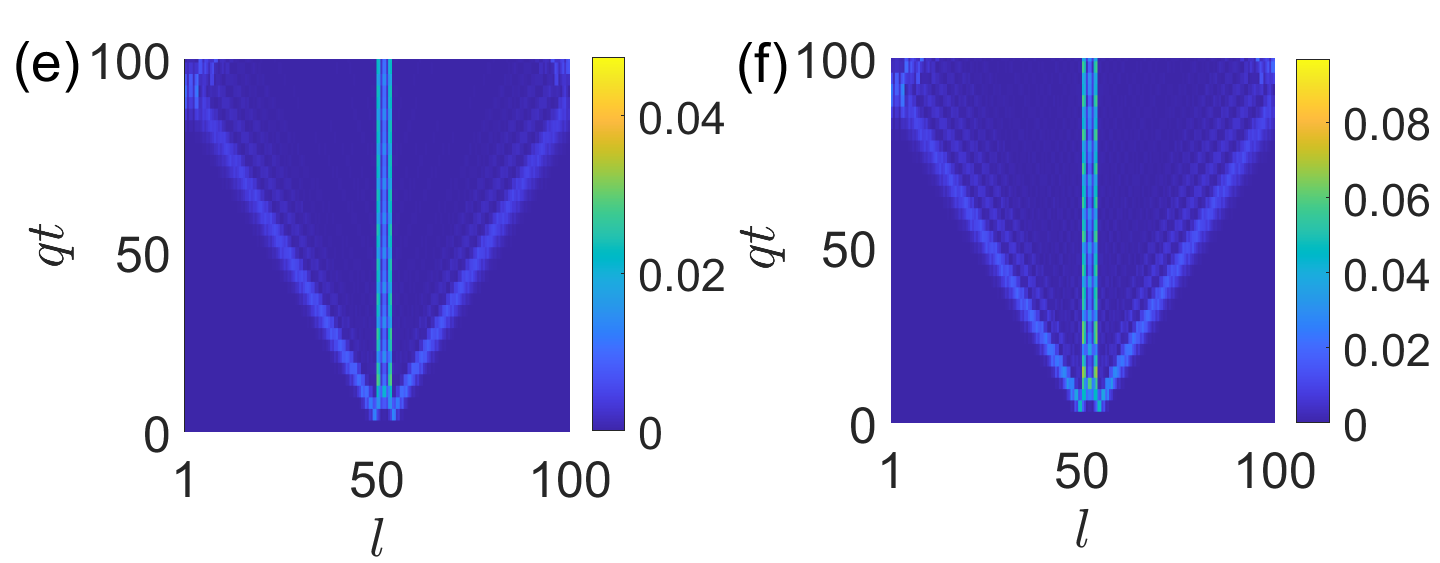}
  \nonumber\\
  \caption{The evolution of the average photon number $n_{il}$ in the SSH chain  for the A-B coupling. The parameters are set as $\delta=0.5, L=100, \theta=0.8\pi, n=50, m=54$.
(a) $(\Omega-\omega)/q=1.99, g/q=0.1$. (b)  $(\Omega-\omega)/q=1.99, g/q=0.2$.
(c) $(\Omega-\omega)/q=1.66, g/q=0.1$. (d)  $(\Omega-\omega)/q=1.66, g/q=0.2$.
(e) $(\Omega-\omega)/q=1.3, g/q=0.1$. (f)  $(\Omega-\omega)/q=1.3, g/q=0.2$.}
\label{photonicdynamics}
\end{figure}

{For the A-B coupling, when the frequency of the giant atom is nearby the upper edge of the band of the waveguide, it is near resonant to a few photon modes with $k\approx0$, and the atom-waveguide interaction Hamiltonian Eq.~(\ref{ABin}) is approximated as
 \begin{eqnarray}
V^{I}_{\rm AB}(t)
&\approx&g[\sigma^{+}|G\rangle\langle \varepsilon|+{\rm H.c.}].
\end{eqnarray}
where
\begin{equation}
|\varepsilon\rangle=\sum_{k\in[-\varepsilon,+\varepsilon]}\frac{1}{\sqrt{L/2}}|E_{k+}\rangle
\end{equation}
and $\varepsilon\rightarrow0$.
It immediately leads to a Rabi oscillation given by Fig.~\ref{markovian2}(b) in Sec.~\ref{Non-Markovian retardation effect} of the main text. In Figs.~\ref{photonicdynamics}(a) and (b), we plot the dynamics of the average photon number $n_{i,l}=\langle C_{i,l}^\dagger C_{i,l}\rangle$ ($i=A,B$) for each site in this case. It shows that, a large portion of the photon is trapped in the range between two coupling
points. Together with Fig.~\ref{markovian2}(b) in Sec.~\ref{Non-Markovian retardation effect}, we can find that the photon and the atom exchange their excitations via a Rabi oscillation whose frequency is proportional to the atom-waveguide coupling strength $g$.}

{Furthermore, corresponding to the parameters in Fig.~\ref{markovian2}(c) of Sec.~\ref{Non-Markovian retardation effect}, the average photon number is given in Figs.~\ref{photonicdynamics}(c) and (d), it clearly shows that the emitted photon will travel along the whole waveguide, therefore, the retardation effect is induced by the finite size of the waveguide under periodical boundary condition.}

{On contrary, for the giant atom induced retardation effect in Fig.~\ref{markovian2}(d) of Sec.~\ref{Non-Markovian retardation effect}, we demonstrate the corresponding average photon number in Figs.~\ref{photonicdynamics}(e) and (f). It shows that the emitted photon in the left (right) atom-waveguide coupling site will cross the regime covered by the atom and travel to the  right (left) coupling site, and exchange excitation with the atom in these two coupling sites. The exchange frequency is independent of the atom-waveguide coupling strength $g$, which is very different from that shown in  Figs.~\ref{photonicdynamics}(a) and (b).}

{Physically speaking, the retardation effect induced by the giant atom can be observed under the following condition (i) the giant atom has a small decay rate so that the lifetime of the atom is much larger than the propagating time of the emitted photon between the atom-waveguide coupling points. (ii) The frequency of the atom is far away from the edge of the energy band of the waveguide.}

\end{document}